\documentclass[longauth]{aa} % for the long lists of affiliations 
\usepackage{graphicx}
\usepackage{natbib}
\usepackage{txfonts}
\begin{document}
   \title{The VVDS-VLA Deep Field:}
   \subtitle{IV: Radio-optical properties}

\author{S.Bardelli \inst{1}
\and E.Zucca    \inst{1}
\and M.Bolzonella  \inst{1} 
\and P.Ciliegi  \inst{1}
\and L.Gregorini \inst{2,6}
\and G.Zamorani \inst{1} 
\and M.Bondi \inst{2}
\and A.Zanichelli \inst{2}
\and L.Tresse \inst{3}
\and D.Vergani \inst{1}
\and I.Gavignaud \inst{5}
\and A.Bongiorno \inst{6}
\and D.Bottini \inst{4}
\and B.Garilli \inst{4}
\and V.Le~Brun \inst{3}
\and O.Le~F\`evre \inst{3}
\and D.Maccagni \inst{4}
\and R.Scaramella \inst{2,8}
\and M.Scodeggio \inst{4}
\and G.Vettolani \inst{2}
\and C.Adami \inst{3}
\and S.Arnouts \inst{3}
\and A.Cappi    \inst{1}
\and S.Charlot \inst{9,10}
\and T.Contini \inst{7}
\and S.Foucaud \inst{11}
\and P.Franzetti \inst{4}
\and L.Guzzo \inst{12}
\and O.Ilbert \inst{13}
\and A.Iovino \inst{12}
\and F.Lamareille \inst{1,3}
\and H.J.McCracken \inst{10,14}
\and B.Marano     \inst{6}  
\and C.Marinoni \inst{15}
\and A.Mazure \inst{3}
\and B.Meneux \inst{4,12}
\and R.Merighi   \inst{1} 
\and S.Paltani \inst{17,18}
\and R.Pell\`o \inst{7}
\and A.Pollo \inst{19}
\and L.Pozzetti    \inst{1} 
\and M.Radovich \inst{20}
\and U.Abbas \inst{3}
\and J.Brinchmann \inst{21}
\and O.Cucciati \inst{12,22}
\and S.de~la~Torre \inst{3}
\and L.de~Ravel \inst{3}
\and P.Memeo \inst{4}
\and E.Perez-Montero \inst{7}
\and Y.Mellier \inst{10,14}
\and P.Merluzzi \inst{20}
\and S.Temporin \inst{12}
%\and C.J.Walcher \inst{3}
\and H.R. De Ruiter  \inst{1}
\and P.Parma  \inst{2}
         }
\offprints{S.Bardelli sandro.bardelli@oabo.inaf.it}  
%\email{sandro.bardelli@oabo.inaf.it}
   \institute{
INAF-Osservatorio Astronomico di Bologna - Via Ranzani,1, I-40127, Bologna, Italy  
\and 
IRA-INAF - Via Gobetti,101, I-40129, Bologna, Italy 
\and
Laboratoire d'Astrophysique de Marseille (UMR6110), CNRS-Universite de
Provence, 38 rue Frederic Joliot-Curie, F-13388 Marseille Cedex 13
\and
IASF-INAF - via Bassini 15, I-20133, Milano, Italy
\and
Astrophysical Institute Potsdam, An der Sternwarte 16, D-14482
Potsdam, Germany
\and
Universit\`a di Bologna, Dipartimento di Astronomia - Via Ranzani,1,
I-40127, Bologna, Italy
\and
Laboratoire d'Astrophysique de Toulouse/Tabres (UMR5572), CNRS, 14 
av. E. Belin, F-31400 Toulouse (France)
%Universit\'e Paul Sabatier - Toulouse III, Observatoire Midi-Pyr\'en\'ee
\and
INAF-Osservatorio Astronomico di Roma - Via di Frascati 33,
I-00040, Monte Porzio Catone,
Italy
\and
Max Planck Institut fur Astrophysik, 85741, Garching, Germany
\and
Institut d'Astrophysique de Paris, UMR 7095, 98 bis Bvd Arago, 75014
Paris, France
\and	
School of Physics \& Astronomy, University of Nottingham, University Park, Nottingham, NG72RD, UK
\and
INAF-Osservatorio Astronomico di Brera - Via Brera 28, Milan,
Italy
\and
Institute for Astronomy, 2680 Woodlawn Dr., University of Hawaii,
Honolulu, Hawaii, 96822
\and
Observatoire de Paris, LERMA, 61 Avenue de l'Observatoire, 75014 Paris, 
France
\and
Centre de Physique Th\'eorique, UMR 6207 CNRS-Universit\'e de Provence, 
F-13288 Marseille France
\and	
School of Physics \& Astronomy, University of Nottingham, University Park, Nottingham, NG72RD, UK
\and
Integral Science Data Centre, ch. d'\'Ecogia 16, CH-1290 Versoix
\and
Geneva Observatory, ch. des Maillettes 51, CH-1290 Sauverny, Switzerland
\and
The Andrzej Soltan Institute for Nuclear Studies, ul. Hoza 69, 00-681 Warsaw, Poland
\and
INAF-Osservatorio Astronomico di Capodimonte - Via Moiariello 16, I-80131, Napoli,
Italy
\and
Centro de Astrofísica da Universidade do Porto, Rua das Estrelas,
4150-762 Porto, Portugal 
\and
Universit\'a di Milano-Bicocca, Dipartimento di Fisica - 
Piazza delle Scienze, 3, I-20126 Milano, Italy
}
   \date{Received -- -- ----; accepted -- -- ----}

% \abstract{}{}{}{}{} 
% 5 {} token are mandatory
 
  \abstract
  % context heading (optional)
  % {} leave it empty if necessary  
   {}
 % aims heading (mandatory)
   {The availability of  wide angle and deep surveys, both in the optical and the radio band, allows us to explore the evolution 
of radio sources with optical counterparts up to redshift $z\sim 1.1$ in an unbiased way, using not only large numbers of radio sources but also 
well defined control samples of radio quiet objects.} 
  % methods heading (mandatory)
   {We use the 1.4 GHz VIMOS-VLA Deep Survey and the optical VIMOS-VLT Deep Survey and the CFHT Legacy Survey to compare the properties
of radio loud galaxies with respect to the whole population of optical galaxies. The availability of multiband photometry and high quality photometric redshifts  
allows to derive rest frame colors and radio luminosity functions down to a limit of a B rest-frame magnitude of $M_B=-20$. 
Moreover, we derive spectrophotometric types, following  the classification of Zucca et al. (2006), in order to have a priori knowledge of the optical  evolution of different galaxy classes.}
  % results heading (mandatory)
  {Galaxy properties and luminosity functions are estimated up to $z\sim 1$ for radio loud and radio quiet early   and late type galaxies.
 Radio loud late type galaxies show significantly redder colors with respect to radio quiet objects of the same class and this is an effect related to the 
presence of more dust in stronger star forming galaxies.
Moreover, we estimate optical luminosity functions, stellar masses and star formation rate distributions for radio sources and compare
them with those derived for a well defined control sample, finding that the probability for a galaxy to be a radio emitter significantly increases 
at high values of  these parameters. Radio loud early type galaxies show luminosity evolution of their bivariate radio-optical luminosity function, 
due to an evolution in the radio-optical ratio.
 The lack of  evolution of the mass function of radio loud early type galaxies means that no new AGN are formed at redshift $z<1$.
On the contrary, radio loud late type objects show a strong evolution, both  in luminosity and in density, of the radio luminosity function for $z>0.7$. 
This evolution is the direct effect of the strong optical evolution of this class and no significant change with redshift of the radio-optical ratio is required. 
With the knowledge of the radio-optical ratio and  the optical and radio luminosity functions for late type galaxies, we show that it is possible to estimate 
the star formation history of the Universe up to redshift $z\sim 1.5$, using optical galaxies as tracers of the global radio emission.}
  % conclusions heading (optional), leave it empty if necessary 
   {}


   \maketitle
%
%________________________________________________________________

\section{Introduction}

Deep 1.4 GHz counts show an upturn below a few millijansky (mJy), corresponding
to a rapid increase in the number of faint sources \citep{Windhorst}. 
First results of the spectroscopic follow-up of relatively bright counterparts ($B<22$) identified 
many of them 
as blue galaxies with spectra showing evidence of intense star formation \citep{Franceschini,Benn}.
But with  increasing depth of the optical follow-up, it resulted that an increasing number of
sub-mJy sources was associated with earlier type galaxies \citep{Gruppioni,Prandoni,Hammer,Afonso06}. 

The relative fractions of the various populations responsible of the sub-mJy radio 
counts (AGN, starburst, late and early type galaxies) are still amatter of debate, with different surveys
providing different results \citep[][and references therein]{Smolcic}. 

In fact, the photometric and spectroscopic work needed in the optical identification process is very 
demanding in terms of telescope time, since a significant fraction of faint radio sources has also 
very faint optical counterparts.  
It is therefore clear that in order to investigate the nature and evolution of the sub-mJy 
population it is necessary to couple deep radio and optical (both imaging 
and spectroscopic) observations over a reasonably large area of the sky. 

The properties of the radio sources in the local Universe have been extensively 
studied by \citet{Best} using the NVSS/FIRST radio surveys coupled with the optical 
Sloan Digital Sky Survey and by \cite{Magliocchetti} using the same radio data and 
the 2dF  galaxy redshift survey for the optical part.  
These optical surveys permit a spectral classification for the optical counterparts 
of radio sources, mainly dividing the sample in early and late type galaxies (broadly 
representative of passive  and star-forming galaxies) on the basis of spectral features such
as the 4000 $\AA$ break or colors. The results, represented by the bivariate radio-optical
luminosity functions, show that the luminosity function of early type galaxies (presumably 
hosting an AGN) can be parametrized with a power law function flatter than that of late 
type galaxies.

Moving to higher redshift, the Phoenix Deep survey allowed the study of the evolution of the 
radio luminosity function \citep{Afonso05} and showed that the radio emission of star-forming 
galaxies strongly evolved from $z\sim 0.1$ up to $z\sim 0.5$.
 More recent surveys explored the star formation density evolution at even higher redshifts
\citep{Smolcic,Ivison,Seymour,Barger07} finding the typical trend of rapidly increasing star formation 
from redshift zero to $\sim 0.8-1.0$ and then a plateau (or a decrease) for higher redshifts.

When one wants to study radio sources as a function of  redshift, it is necessary to go 
through the radio-optical identification process. For this reason, it is critical to know 
as accurately as possible the properties of the optical counterparts 
in order to discriminate which is an effect due to the optical or to  radio evolution. 

The VIMOS-VLT Deep Survey \citep{Lefevre2005a}, complemented with  photometric data 
from the CFHT Legacy Survey, gives a unique opportunity to study the evolution of optical 
galaxies. In fact, it offers the availability of a large sample of reference redshifts
obtained from a deep spectroscopic survey, a large set of photometric bands and high quality 
photometric redshifts over a wide area.

This allows us to study the properties of the counterparts of an associated 1.4 GHz VLA survey 
\citep{Bondi,Ciliegi1} using the rest-frame colors and therefore avoiding the use 
of redshift dependent quantities.  

 The aim of this paper is to study the optical properties of  galaxies hosting a radio source 
with respect to a well defined control sample of galaxies, for which the selection effects and the 
evolutionary behaviour are known a priori, in order to understand the
relations between the radio and  optical properties.
For this reason, in this paper we define as ``radio galaxies" or ``radio loud galaxies" 
all galaxies with  detected radio emission: therefore our definition is based only
on observed quantities. 

 The advantage of this work is the simultaneous availability  of a deep radio survey, a multiwavelength 
optical band coverage over a relatively large area, accurate photometric redshifts 
and absolute magnitudes which minimize the model dependency,
but the most important point is the availability 
of a control sample with well studied luminosity and stellar mass  functions as a function  
of redshift and galaxy type, from which it is possible to extract in a completely homogeneus way the galaxies 
which host a radio source.

The plan of the paper is the following: in Section 2 we present the data and the definition 
of our samples, in Section 3 we compare the redshift distribution of the radio galaxies subsample with the sample of all galaxies, 
in Section 4 we present the results of the color distributions 
and in Section 5 the bivariate luminosity functions. 
The radio-optical ratio distribution for our sample is presented in Section 6,
while the stellar mass and star formation rate distributions are shown in Section 7.
We estimate in Section 8 the star formation density evolution and in Section 9 we present the conclusions.

Throughout the paper we adopt a flat $\Omega_m = 0.3$ and
$\Omega_\Lambda = 0.7$ cosmology, with $H_o = 70$ km s$^{-1}$ Mpc$^{-1}$. 
Absolute magnitudes are given in the AB system and are expressed in
the five standard bands U (Bessel), B and V (Johnson), R and I
(Cousins).
Errors are estimated as poissonian fluctuation and refer to $1\sigma$ confidence
level.

%---------------------------------------------------------------------------------------
\section{The data}

%_____________________________________________________________
%
\begin{table*}
\centering          
\label{table:1}     
\caption{Numbers of objects in our samples. 
The low/high redshift bins are [0-0.5]/[0.5-1.1] for all galaxies and type 3+4 galaxies 
samples and [0-0.7]/[0.7-1.0] for type 1+2 galaxies sample. }  
\begin{tabular}{c c c c c c c c c}     % 7 columns 
\hline     
\smallskip
redshift interval &\multicolumn{4}{c}{Complete Sample} &  \multicolumn{4}{c}{Control Sample} \\
\hline
 \smallskip
   &  all & type 1 ($z<1.0$)  & type 1+2 & type 3+4 &  all & type 1 ($z<1.0$) & type 1+2 & type 3+4    \\ 
\hline                    
\smallskip
 $0<z<1.1 $ & 430 & 149 & 303 & 127 & 17880 & 3775 & 7414 & 10466 \\ 
\smallskip
low z & 238 &  & 176 & 62 & & & & \\
\smallskip
high z & 150 & & 91 & 59 & & & & \\
\hline                  
\end{tabular}
\label{numtab}
\end{table*}
%
%_____________________________________________________________
%                                          Table with foonotes 

The radio data were obtained with the Very Large Array (VLA)
in B configuration: a one square degree mosaic map with an 
approximately uniform noise of $\sim 17.5 \mu$Jy (1$\sigma$) and 
with a 6  $\times$ 6 arcsec FWHM gaussian resolution beam has 
been obtained. 
This map (centered at $\alpha_{2000}$ = 02:26:00 $\delta_{2000}$ = $-$04:30:00)
has been used to extract a  catalogue of 1054 radio sources complete 
to a limit of $80 \mu$Jy.
 A detailed description of the radio observations, data reduction,
sources extraction and radio source counts is reported in \cite{Bondi}, while
the radio--optical correlation  (done with a likelihood ratio technique) and the properties of radio objects with optical 
counterparts are presented in \cite{Ciliegi1}, who found 718 radio sources 
with an optical counterpart brighter than $I_{AB}=25$.
The percentage of optical identifications of the radio sources varies from $\sim 70\%$, 
for fluxes between 0.08 and 0.2 mJy, to $\sim 65\%$, for fluxes between 0.2 and 0.1 mJy,
and decreases to $58\%$ for brighter sources.
Moreover, \cite{Bondigmrt} performed a 610 MHz survey with the Giant Meterwave Radio Telescope
(GMRT), finding that the median spectral index of faint radio sources 
(below 0.5 mJy at 1.4 GHz) is significantly flatter than that of brighter sources. 
This fact is probably due to a relevant contribution below 
0.5 mJy from a population of flat spectrum low luminosity compact AGN and radio quiet QSO.

The optical data are part of both the F02 deep field of the VIMOS-VLT Deep Survey 
(VVDS) and the D1 field of the MEGACAM CFHT Legacy Survey (CFHT-LS).
The first set comprises the B, V, R and I bands \citep{McCrackenVVDS}, 
obtained with the CFHT wide-field 12 K mosaic camera on the entire field.  The completeness 
limits are $\sim 26 $ for the B, V and R bands and $\sim 25$ for the I band.

The second set comprises the $u^*, g',r',i',z'$ bands \citep{McCrackenLegacy} and  
covers an area of 1 deg$^2$ centered on the VVDS F02 deep field;  the completeness limits are
26.0, 25.8, 24.3, 25.1, 24.5, respectively.
Moreover, we used the UKIDSS J and K data \citep[ limited at K$\sim 18$,][]{UKIDSS}, covering the entire area.

Starting from the VVDS photometric catalogue, a deep spectroscopic survey has been conducted for
galaxies with I$_{AB}\le$24 \citep{Lefevre2005a}, with a sampling rate of $\sim 33\%$ over 
an area of $0.6$ deg$^2$  and with a spectral resolution of $\sim 33 \AA$ at $7500 \AA$.  
Due to this sampling rate and the area covered, only 53 radio sources (among the 718 with optical
counterpart) have a spectroscopic redshift. Therefore, in order to increase the statistics,
we decided to use the VVDS photometric redshift sample.
The VVDS photometric redshift catalogue, the method to estimate redshifts and their quality are presented in  
\cite{Ilbertphz}. The typical error is $\sigma(\Delta z / (1+z))\sim 0.029$ and the reliability is maximum 
in the redshift range $[0.2-1.5]$.
 We analysed the differences between the photometric and spectroscopic redshifts 
for the 28 galaxies in the Radio Sample (see below), finding that the zero point is 
consistent with zero (at $0.05 \sigma$) and the error is 0.034. Therefore the properties of the photometric
redshifts of radio galaxies are consistent with those of the optical sample.

Absolute magnitudes are computed following the method described in the
Appendix of \cite{Ilbertlf}: the K-correction is computed
using a set of templates and all the photometric information 
available. However, in order to reduce the template dependency, the
rest frame absolute magnitude in each band is derived using the
apparent magnitude from the closest observed band, shifted at the
redshift of the galaxy. With this method, the applied K-correction is
as small as possible.

Following the procedure adopted in \cite{Zucca}, 
for each galaxy the rest frame magnitudes were matched with an 
empirical set of SEDs, composed of four observed spectra \citep[CWW][]{CWW} and
two starburst SEDs computed with GISSEL \citep{bc93}.  
The match is performed minimizing a $\chi^2$ variable on these templates at the 
redshift of each galaxy. 
 Note that this method for the galaxy classification is more refined than 
a simple color division because of the siultaneous use of all the available photometric 
data, which minimizes the influence of errors in a single band.

Galaxies have been divided in four types, corresponding to the E/S0
template (type 1), early spiral template (type 2), late spiral template
(type 3) and irregular template (type 4); type 4 includes also the 
starburst templates. 
\cite{Franzetti} have shown that there is a good correspondence between this classification 
and that obtained from  spectral features, like e.g. the 4000 $\AA$ break up to $z\sim 1.4$.  

As shown by \cite{Zucca} studying the luminosity function of these  photometric types for the spectroscopic survey, 
 type 1 (corresponding to early type galaxies) and type 4 galaxies (corresponding to star forming late blue galaxies)
show the most extreme behaviour regarding the evolution of the luminosity function, with the first class showing little evolution and the latter showing a strong evolution.
The other two classes (2 and 3) could be regarded as intermediate types in the sense that their properties are less extreme. 

The accurate knowledge of the behaviour of these classes led us to use the same classification: here, we assume that the results 
for the spectroscopic subsample holds also for the entire, photometric redshift based sample. This assumption is confirmed
by the comparison of the luminosity functions from the two datasets (O.Ilbert, private communication).

From the 718 original radio-optical identifications (the ``Original Sample''), we  consider only  objects with 
$M_{B}<-20$  in 
the redshift interval [0-1.1] ([0-1.0] when considering type 1 alone), where we have reliable photometric redshifts and well studied optical luminosity functions. 
We will refer to this sub-sample as the ``Complete Sample''.
The magnitude limit assures that the considered galaxies are visible within the full adopted redshift range and 
therefore our sample has the same properties as that of a volume limited sample. 
The different redshift limit for type 1 galaxies is due their higher  K-correction, which influences their absolute magnitude limit  \citep[see the discussion of][]{Zucca}.

\begin{figure}
   \centering
   \includegraphics[width=\hsize]{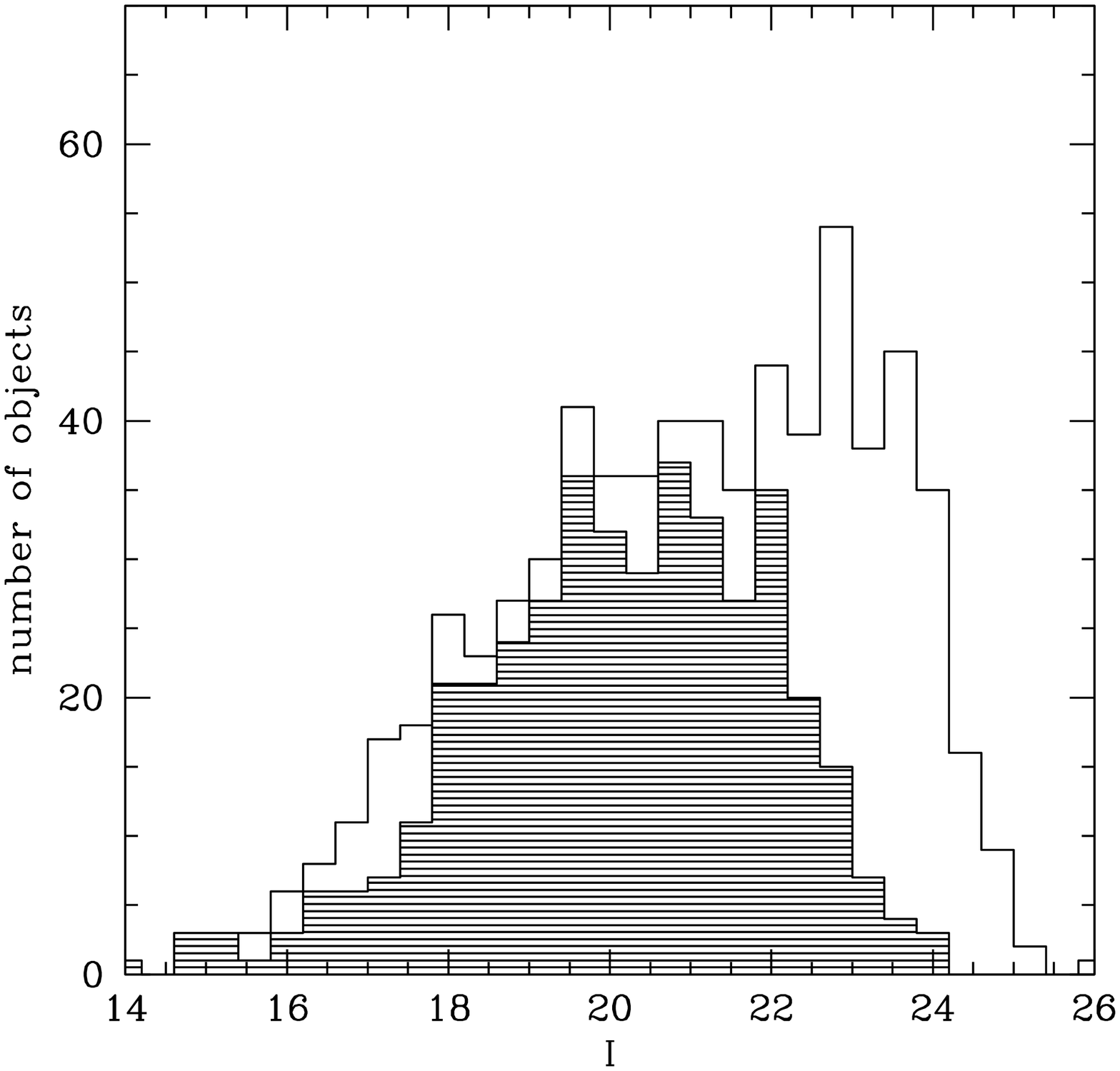}
   \includegraphics[width=\hsize]{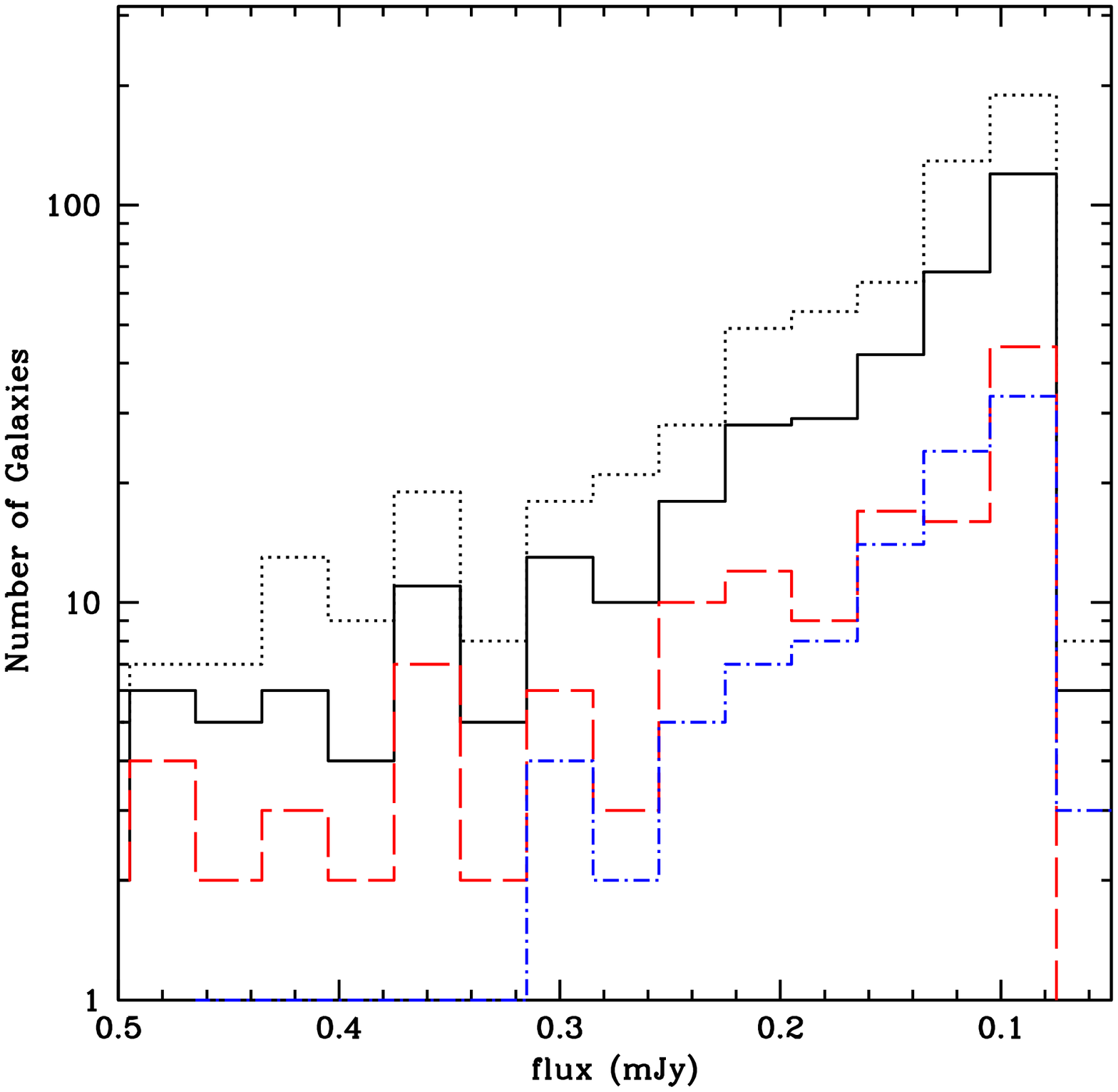}	
      \caption{ 
{\it Upper panel:} Shaded histogram: distribution of I magnitude for radio galaxies with $M_{B}<-20$
	and $z_{phot}<1.1$ (the Complete Sample). Empty histogram: all identified radio galaxies (the Original Sample). 
{\it Lower Panel:}
Dotted histogram corresponds to the radio flux for all radio galaxies (the Original Sample), solid line represents the Complete Sample, dot-dashed 
line the type 3+4 objects,  long dashed line the type 1 objects.}
         \label{histo_magflux}
   \end{figure}

The sample of all galaxies from the CFHT-LS/VVDS catalogue has been cut at the same limits obtaining in this way 
the ``Control Sample''. 
In the upper panel of Figure \ref{histo_magflux} we show the apparent I-band magnitude distribution of the Original Sample (empty histogram) and of the 
Complete Sample (shaded histogram). This histogram allows us to define a further cut in the 
Control and Complete Samples at I$_{AB}<$24. This makes the two samples cleaner because at fainter magnitudes the 
estimated photometric redshifts are less precise and the fraction of wrong redshift increases.
 Moreover, this limit in apparent magnitude is the same as for VVDS spectroscopic survey and therefore
we are in the same framework as  \cite{Zucca} work on the luminosity functions. 

Within these limits there are 430 radio sources.  It is clear from the figure that the majority of the objects not included in the Complete Sample have magnitudes
 I$_{AB}>$22. Among the radio sources not selected by our limits,
133 have  $M_{B}>-20$ and 154 match our absolute magnitude criterion but have $z>1.1$, while the 35 I$_{AB}$$<18$ objects 
are  underluminous to be included in the Complete Sample.

Considering the types distribution within the Complete Sample, 169 objects are type 1 (149 within $z<1.0$), 134 are type 2, while 74 and 53 are type 3 and
 type 4, respectively.  Given the relatively low number of type 4 galaxies, we will consider as a single class (hereafter type 3+4) the
late type galaxies. The Control Sample is formed by 4332 type 1 (3775 $z<1.0$), 3082 type 2 and 10466 type 3+4 objects
(see Table \ref{numtab}). Therefore in our sample the percentage of 
radio emitting sources is $\sim 4 \%$, $\sim 3\%$ 
and $\sim 1\%$ for the three considered classes, respectively. 
Here we assume that the emission mechanism of type 1 and type 2 radio galaxies is
mainly due to AGN  \citep[thus selecting the ``absorption line AGN" of ][]{Best}, while for type 3 and 
type 4 it is induced by star formation.

  This assumption is reasonable in the light of the classification in Figure 9 of 
\cite{Best}, which shows the plot log$L_{radio}/M_*$ versus $D_n (4000)$, where $L_{radio}$ is
the radio luminosity, $M_*$ is the stellar mass and $D_n (4000)$ is the measure of the
$4000\AA$ break. 
In this plot there are two well separated clumps, not depending on log$L_{radio}/M_*$, 
one at $D_n (4000)\sim 1.25$ populated almost completely by star forming galaxies, and 
the other at $D_n (4000)\sim 2.0$ dominated by galaxies with AGN induced emission and LINERS.
For the 31 radio galaxies in our sample with measured  $D_n (4000)$ our classification agrees
with that of \cite{Best}.
Considering the whole sample, our radio sources have almost all ($\sim 93\%$) a ratio 
$11.5<$log$L_{radio}/M_* < 14 $ and this implies that the division between star forming and 
AGN induced radio emission spans the range [1.3 - 1.6] in $D_n (4000)$: this value
roughly corresponds to the division between type 1+2 and type 3+4 for the Control Sample. 

Another way to test our classification by looking at the upper panel of Figure \ref{lumtosfr} 
(see Section 7): all radio galaxies with star formation rates (as estimated in the optical) less than 
$1-2 M_{\odot} yrs^{-1}$ (all of type 1+2) are far from the correlation between SFR(optical) and 
SFR(radio) seen at higher rates, meaning that the AGN component of the radio luminosity  is dominant.
Moreover, we checked our classification with the spectra of the zCOSMOS 10K sample \citep{zcosmos}, 
by limiting the data set at $M_B<-20$ and $z<0.9$, due to the different depth of the survey. 
It results that only $6 \%$ of the type 1+2 objects falls under the dividing line (i.e. classified as star forming)  
of \cite{Best} and $30 \%$ of type 3+4 are just above the dividing line \citep{Bardelli}. 

In the lower panel of Figure \ref{histo_magflux} we show the 1.4 GHz flux  histogram of the Original Sample and the 
Complete Sample, while the long dashed line corresponds to type 1 galaxies and the dot-dashed line corresponds to type 3+4 objects.  
There is no major trend of types with flux: the fraction of type 1 galaxies decreases  from $46\%$ in the flux 
range [0.2-0.45] mJy to $32\%$ in the range [0.08-0.2], while type 3+4 galaxies increases
from  $22\%$ to  $31\%$.
 In our case, no dramatic increasing percentage of star-forming galaxies is present at low fluxes, differently from early results \citep[see the discussion in][]{Gruppioni99}.
This conclusion would be similar if we adopt a fainter I magnitude limit or if we relax  all our restrictions, in particular that at  high redshift: 
in this latter case type 3+4  increases from  $37\%$ to $41\%$. Only the type 1 show a decrease from $34\%$ to $25 \%$,
 possibly due to the blueing of ellipticals at higher redshifts, which could shift radio galaxies from type 1 to type 2.
 However, as shown by \cite{Seymour}, our adopted flux limit is just above the upturn of the star forming galaxies counts.

\subsection{Contribution from optically identified AGN}

In the optical spectroscopy, the emission of an active nucleus is revealed by the presence of typical features of  emission lines, leading to the definition of 
two broad classes: the Broad Line AGN (BLAGN) and Narrow Line AGN (NLAGN). These AGN
are known to be sometimes radio loud. 
 These objects are expected to be relatively rare within our limits 
and have spectral energy distributions which could be misclassified
within our classification scheme (AGN induced radio emission associated 
to early type galaxies and star formation induced emission
to late type galaxies). For this reason we consider as ``contamination" 
the presence of these objects in both the Complete and Control samples.

Here, we estimate how large the possible contamination in our sample by these objects is:
unfortunately, a direct estimation is not possible because of the lack of spectroscopy for most of our radio sample. 
   
The presence of optical AGN in the spectroscopic VVDS survey has been studied by \cite{Bongiorno}, \cite{Gavignaud} and \cite{Paltani}, finding a 
sample of 333 NLAGN and 63 BLAGN. 
Within our limits in redshift and absolute magnitude the ratio between 
type 1+2 objects versus type 3+4 is 1/6 for BLAGN and 1/3 for NLAGN.
Among these objects, 4 BLAGN and 2
NLAGN have radio counterparts in the Original Sample and 2 BLAGN and 2 NLAGN are part of our Complete Sample.

The NLAGN optical counterparts have been assigned to the type 1+2 sample, while the BLAGN  to the type 3+4 sample.
The radio luminosities are logP(W Hz$^{-1}$) $=23.19$ and logP(W Hz$^{-1}$)$=22.52$ for NLAGN and
logP(W Hz$^{-1}$)$=23.84$ and logP(W Hz$^{-1}$)$=24.54$ for BLAGN.
These numbers come from the common area between the CFHT-LS  survey and the spectroscopic VVDS survey, 
which is approximately  $28 \%$. 
Moreover we corrected for the VVDS survey sampling rate ($33\%$). Under the assumption that
the contaminaton estimate  obtained in this sub-area is representative of the entire area sampled, we expect in
our sample a number of $21 \pm 15$ AGN objects in the power range logP(W Hz$^{-1}$) =$23-24$.

 As an independent way to estimate the contamination, we correlated the positions of SWIRE survey sources \citep{swire} with our Complete Sample.
For the sample containing type 3+4 we have 85 common galaxies and
a clear relation between redshift and the ratio of fluxes in the 24 $\mu$m and radio bands. Only 9 objects are significantly shifted towards lower 
values of the ratio, indicating the possible presence of an AGN.  If the ratio between all galaxies and AGN is representative, we expect a contamination 
in the sample of $15 \pm 5$ AGN, well consistent with the previous result. This approach is not possible for early type galaxies because the AGN and star forming populations are not separable.

Considering the color distribution of all NLAGN, we find that 
for  type 1 objects the B-I colors are more similar to the radio emitting objects, while the U-B colors have a flatter distribution, with a small peak
 at U-B$\sim 0.7$. Also for type 4 galaxies, colors distributions have a similar behaviour.

%_____________________________________________________________
%                 A figure as large as the width of the column
%-------------------------------------------------------------
%-----------------------------------

\section{The Redshift Distributions }

\begin{figure}
   \centering
   \includegraphics[width=\hsize]{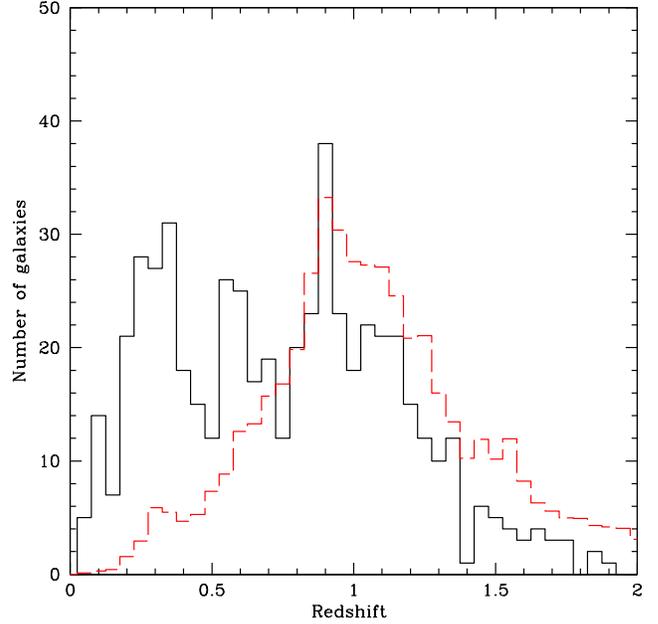}
\caption{Redshift histogram of the radio galaxies in the Complete Sample 
(solid line) compared with galaxies in the Control Sample (dashed line). The Control Sample histogram has been rescaled by the ratio of the numbers of objects in the two samples.}
  \label{histoz}
   \end{figure}

\begin{figure}
   \centering
   \includegraphics[width=\hsize]{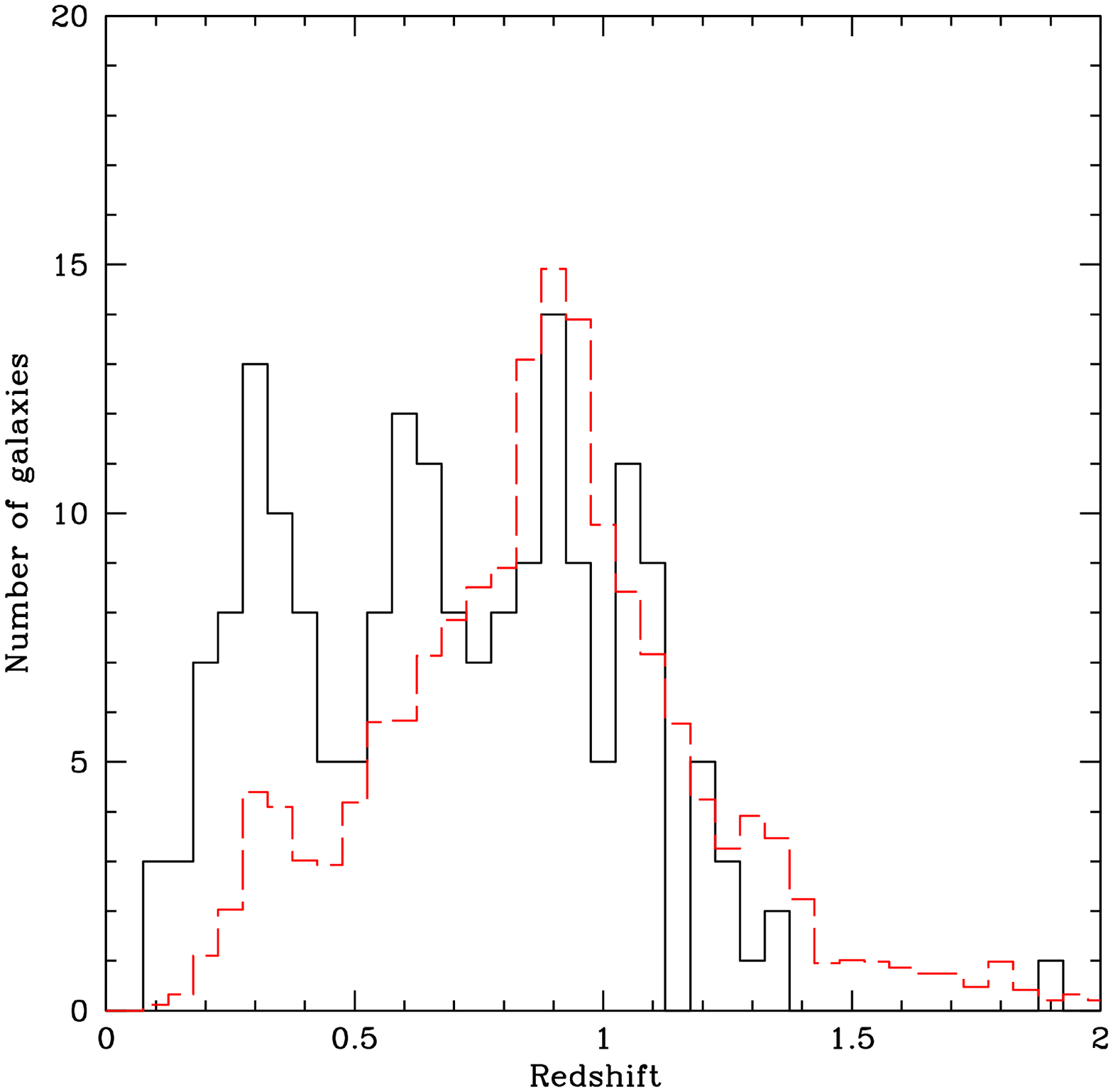}
   \includegraphics[width=\hsize]{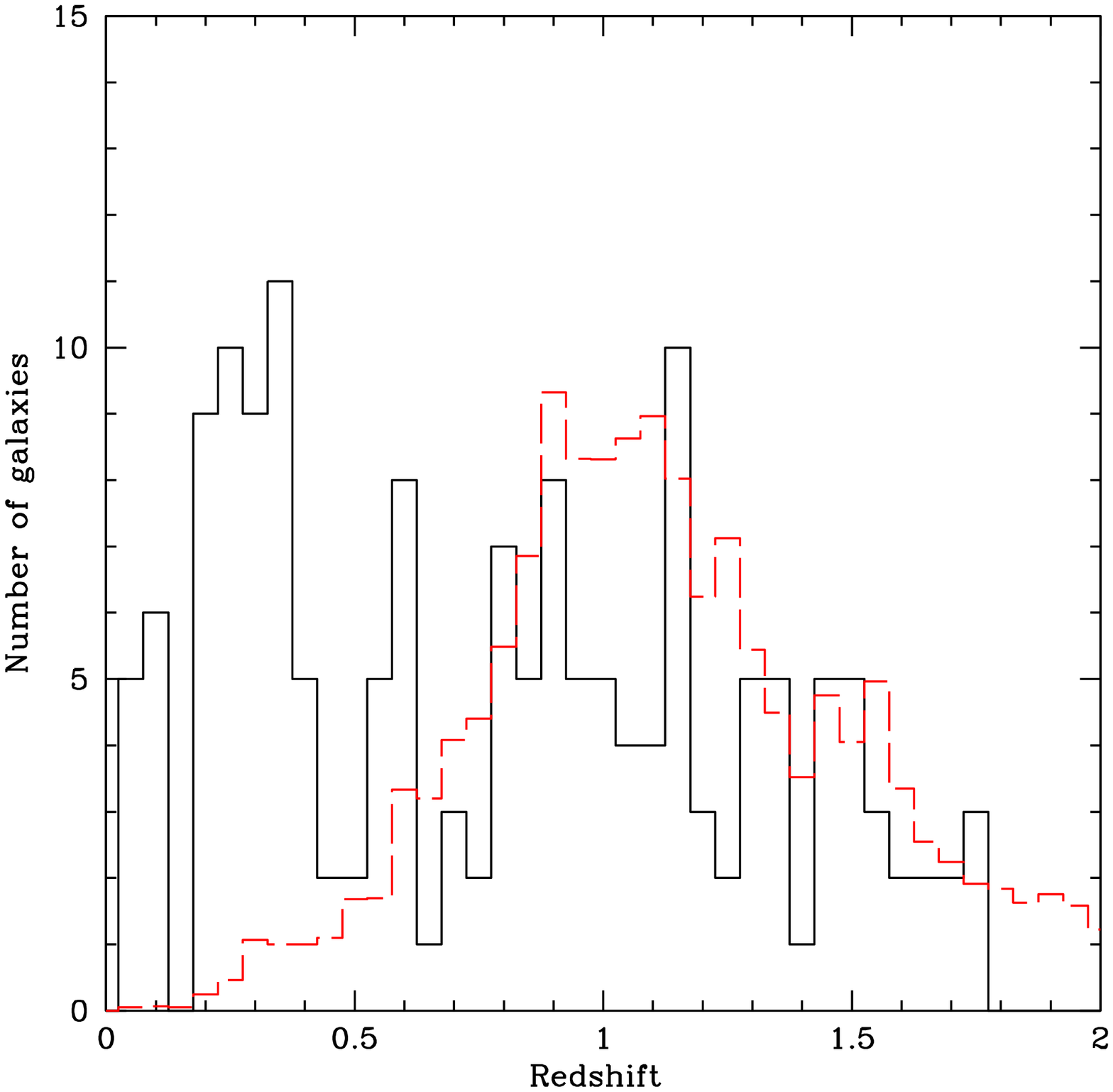}
  \caption{
{\it Upper panel:} Same as Figure \ref{histoz} but for type 1 galaxies.
{\it Lower  panel:} Same as upper panel but for type 3+4 galaxies.
} \label{histozel}
   \end{figure}

In Figures \ref{histoz} and \ref{histozel} the redshift distribution of the radio galaxies of our Complete Sample (solid lines)
is compared 
with the Control Sample (dashed lines) for all galaxy types and for type 1 and 3+4. Although not used in the following analysis, we show in the figures also objects with redshift larger than 1.1.
 Note that the Control Sample histograms have been rescaled by the ratio of the  number of objects in the two samples.
In principle, the two histograms are not expected to be exactly the same, being the radio sample cut both in optical magnitude and in radio flux while the Control Sample has only the first limit.
The most important  peak in both the  Control and in the Complete Sample is at $z\sim 0.9$ and is  mostly due to  early type
galaxies. This maximum
is the result a the combination of the presence of a huge large-scale structure at that redshift  \citep[see][]{Lefevre2005a} and 
the increasing incompleteness of our Complete Sample beyond $z=1.1$ (or $z=1.0$ for type 1 galaxies) induced by our absolute magnitude limit. At lower redshifts, the histogram of 
the Control Sample is determined only by the increasing volume sampled.

The other significant bump visible in the Complete Sample is centered at $z \sim 0.3$ and is mostly due to the radio flux limit.
In fact, within our limits, the radio sample is complete for radio powers 
 of logP(W Hz$^{-1}$)$>$22 and for  $z<0.2$.
At lower redshifts, the distribution is determined only by the increasing volume, while at higher redshifts the histogram is a combination of the losses due to the radio flux limit 
and the increasing numbers of optical galaxies.
 However, also after a visual check, no obvious $\alpha - \delta$ segregation of radio galaxies is present corresponding to this bump.
A  clump at the same redshift is present in the Control Sample and in this case we detected the presence of a 
cluster  (detected as a two-dimensional galaxy distribution)  at $\alpha_{2000}=2^h 24^m 32^s$ and $\delta_{2000}=-4^o 14' 54''$. Two radio sources have redshifts consistent with that of the cluster,
 one being at its centre.

\begin{figure}
   \centering
  \includegraphics[width=\hsize]{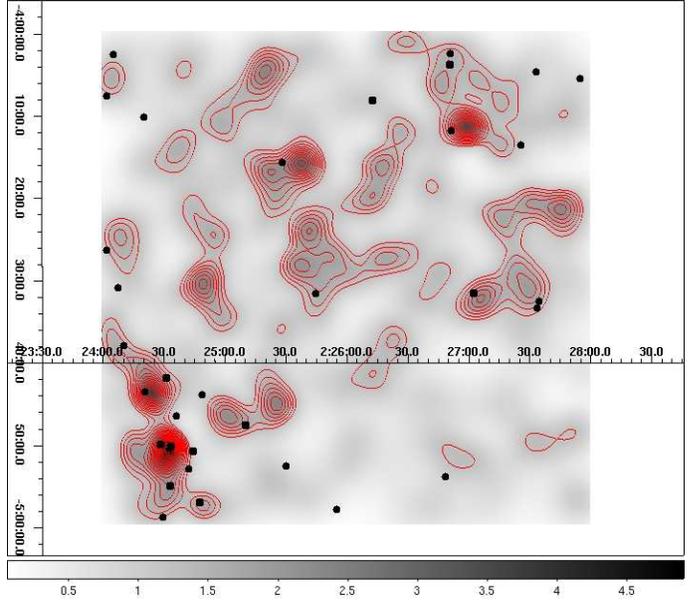}
\caption{Isodensity contours are the adaptive kernel smoothed $\alpha-\delta$ distibution of early type galaxies in the redshift range [0.5-0.65]. 
Points show the positions of the optical counterparts of the radio sources in the same redshift bin.}
  \label{overdens_ealry}
   \end{figure}

 Finally, another difference between Control Samples and Complete Samples is an excess of radio galaxies
 at $z\sim 0.6$, which is the only one
 related to a structure, because of its narrow redshift width.
 By plotting the $\alpha-\delta$ distribution of early type radio galaxies  
(which are expected to better delineate clusters and large scale structures) in the $[0.5-0.65]$ redshift range, it is clear that they are 
mainly in the lower left (South-East) quadrant (Figure \ref{overdens_ealry}).
 By plotting the adaptively smoothed isodensity contours of all galaxies in this bin, we found that the overdensity 
coincides with two galaxy clumps close to each other ($\sim 4$ Mpc). 
This fact is consistent with the claim of 
 \cite{Owen99} that cluster merging triggers radio emission \citep[see also][]{Giacintucci}.
Unfortunately, no velocity dispersion estimate is possible because the two clumps lie just outside the region 
covered with the spectroscopic survey.

 Finally, note that the effect of  cosmic variance on these plots is 
of the order of $15 \%$, as estimated with similar samples,
the same data (CFHT-LS) and the same classification method by \cite{McCracken} 
and more specifically for the VVDS survey by \cite{Vergani}. 
The expected variance on the radio source sample is of the same order because the correlation 
functions in the Control and Radio samples are similar (see Section 3.1).

%-------------------------------

%--------------

%-------------------------------
   \begin{figure}
   \centering
   \includegraphics[width=\hsize]{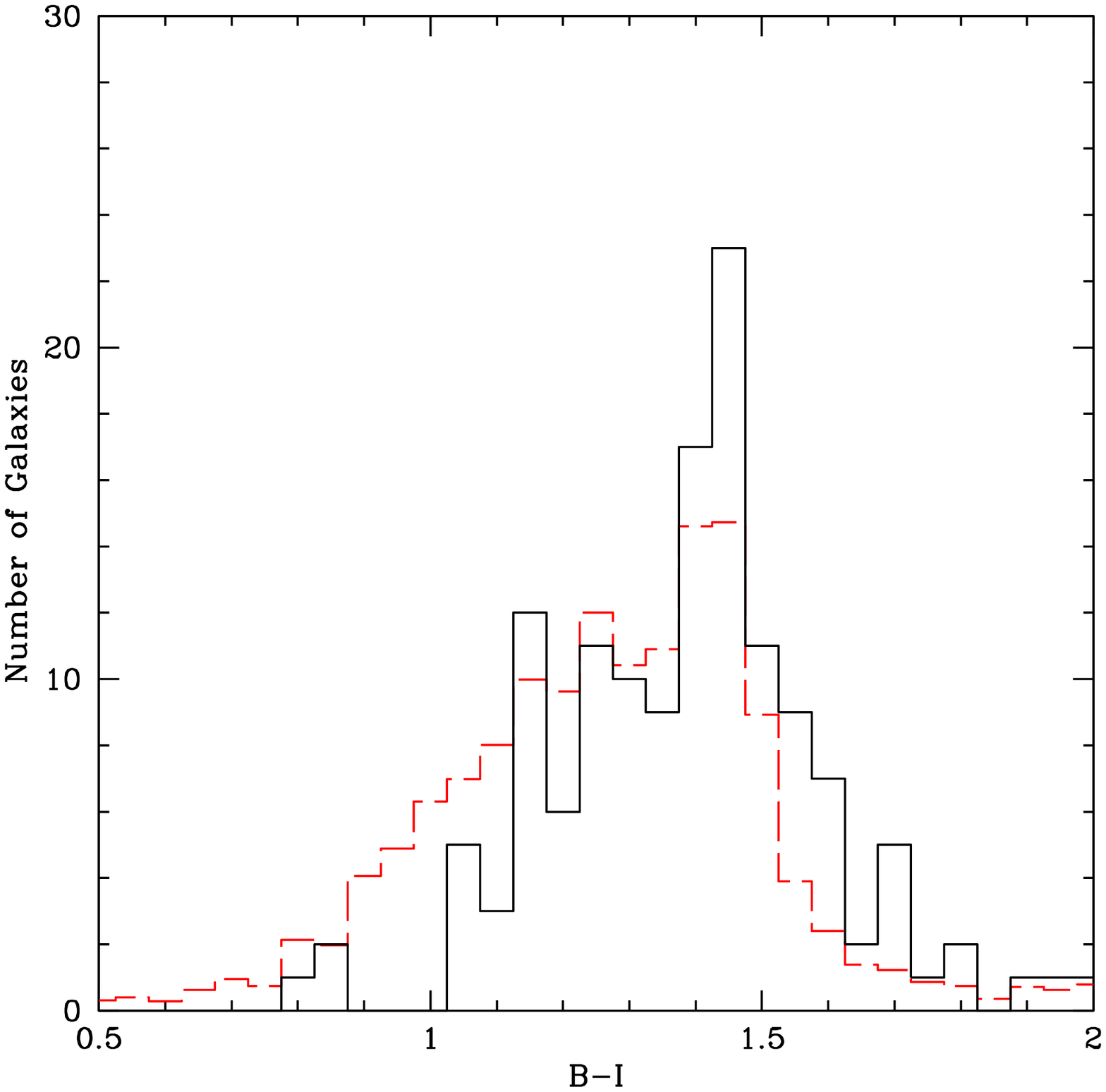}
   \includegraphics[width=\hsize]{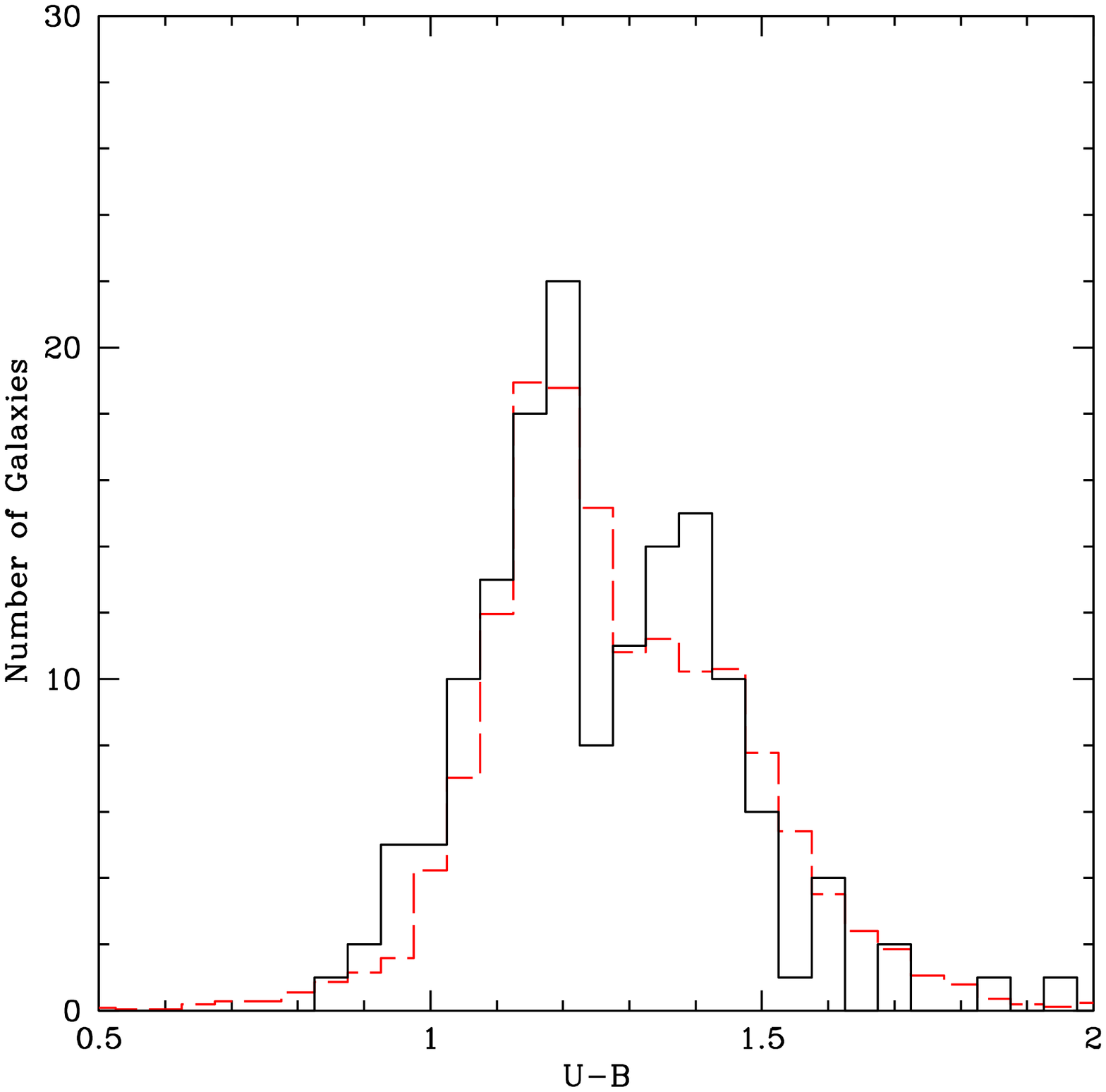}
  \caption{
{\it Upper Panel:} B-I  color distribution of the radio emitting early type galaxies (solid line) compared with the
same distribution of the Control Sample (dashed line). The latter has been rescaled to take into account the different number of objects. 
{\it Lower panel:} U-B color distribution of the radio emitting early type galaxies and of the Control Sample.  Symbols are the same as in the upper panel.
}
\label{bihistorealy_new}
   \end{figure}

%---------------------------------------------
   \begin{figure}
   \centering
    \includegraphics[width=\hsize]{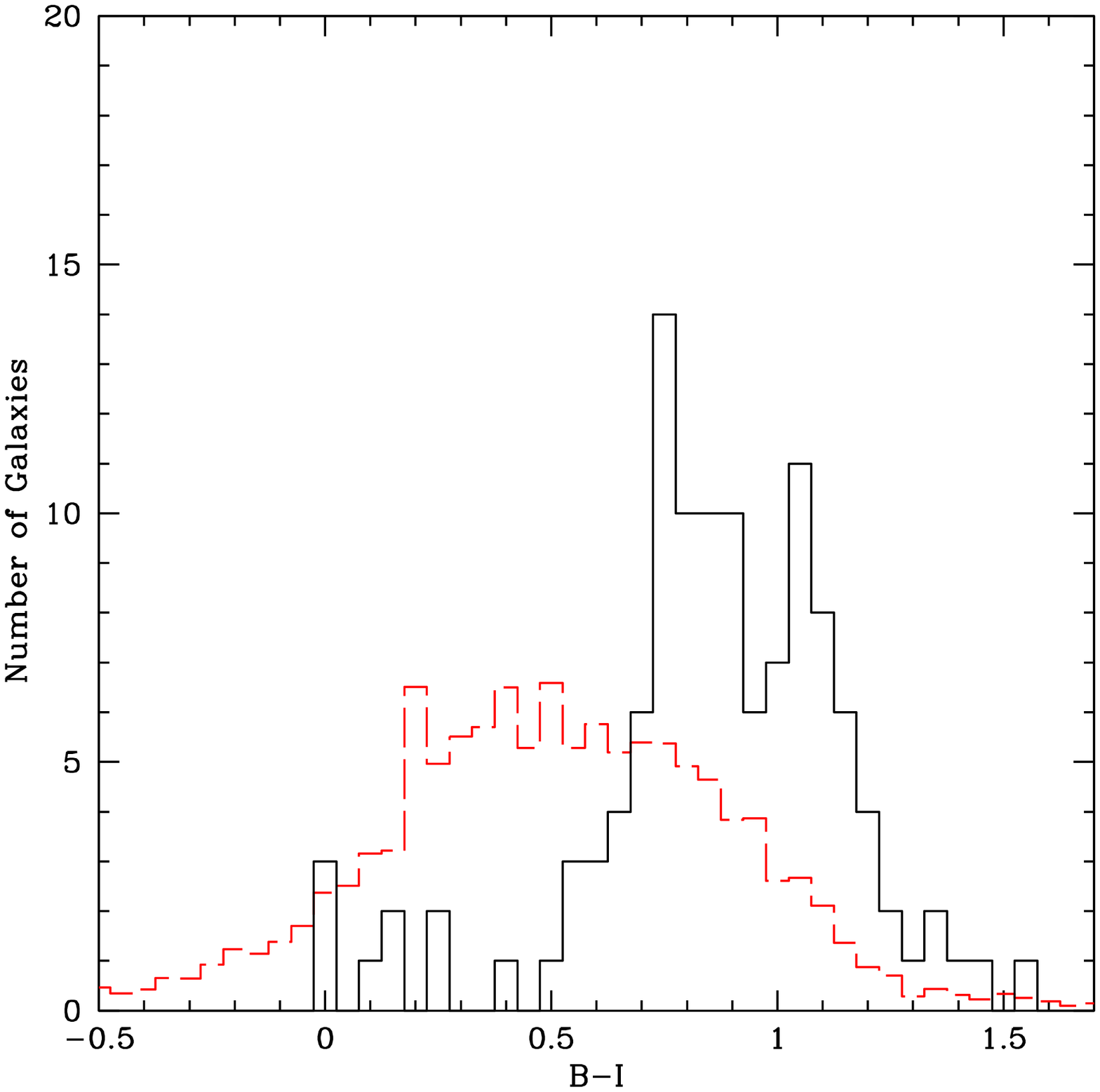}
    \includegraphics[width=\hsize]{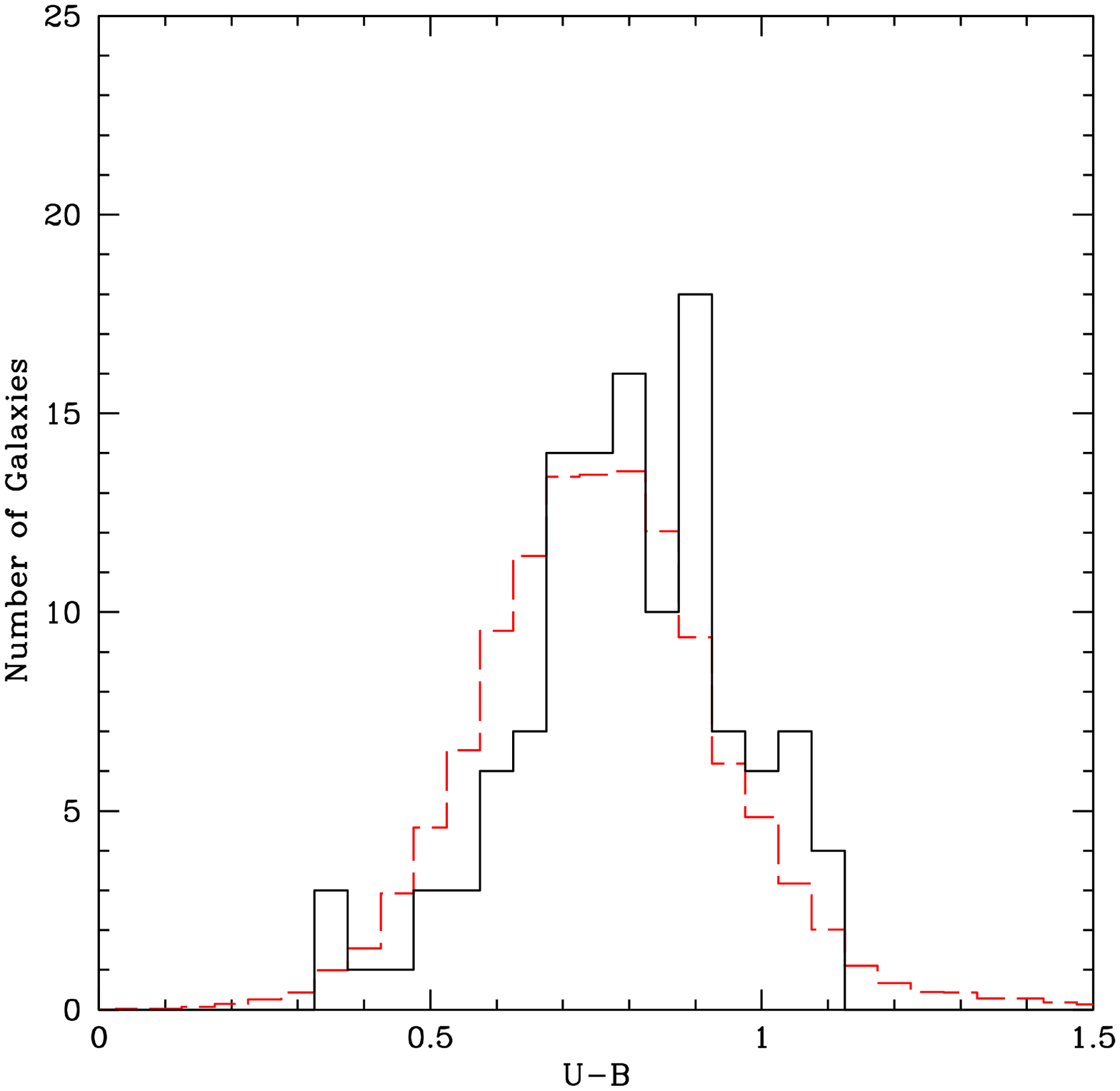}	 
\caption{Same as Figure \ref{bihistorealy_new}  but for type 3+4 objects.}
         \label{ubilate}
   \end{figure}
%--------------------

\subsection{Correlation function}

An open problem concerning the properties of the radio galaxies regards their clustering properties. For this reason we computed the angular correlation function for 
all galaxies and for the radio loud objects divided in type 1 and type 3+4 galaxies.
We did not detect any significant difference among the various populations;
there is only a  weak signal considering the bright and faint type 1 radio galaxies, the ones fainter than logP(W Hz$^{-1}$)$<23.5$ being more clustered than the brighter ones. 

\section{Color Distributions}

The advantage of using redshifts is that one can use the rest-frame colors, by--passing all ambiguities led by the different type dependent K-corrections in adopting observed colors.
Moreover, in order to disentangle real effects due to the presence of the radio source from the evolution of
the optical host galaxy, it is necessary to make a relative comparison between radio galaxies of a given type with their ``parent''  population (in our case the Control Sample).

The rest-frame  B-I color distribution of early type galaxies (type 1) which show radio emission is significantly 
different from that of the parent sample of early type galaxies 
(K-S probability to be extracted from the same population of $\sim 10^{-7}$), the first showing an evident peak at 
B-I $\sim 1.4$ (see Figure \ref{bihistorealy_new} upper panel), the second being broader, with approximately the same mode.
 
The color distribution difference between radio emitting and parent sample early type galaxies is present only at higher redshifts 
($0.5<z<1.0$), while the distributions are statistically indistinguishable for $z<0.5$.
Moreover, the colors of early type powerful radio galaxies (logP(W Hz$^{-1}$)$>23.5$) are not different from those of the   
Control Sample, while  a significant difference  (K-S probability of $\sim 10^{-11}$) is present between the color distribution
of faint sources (logP(W Hz$^{-1}$)$<23.5$) and that of all early type galaxies. 
In all these cases the difference is due to a more peaked and narrower distribution of the radio sources with respect to the 
general population of early type galaxies.
The redshift and luminosity effect are related to each other in the sense that the higher redshift bin is populated mainly by 
the most powerful radio galaxies, while at lower redshifts the dominant population is the faint one. 

The difference in color distribution is present, although marginally significant, also in the U-B distribution
(Figure \ref{bihistorealy_new} lower panel), where the significance of the difference of the radio faint and bright galaxies 
with the Control Sample is $98 \%$ and $95\%$, respectively.

Considering the relative color distribution inside the radio sample, powerful radio sources tend to be bluer in B-I and redder in U-B than faint radio sources.
 This effect  seems to be due to a relatively lower B band flux for radio faint objects.

The difference in the B-I colors between radio loud and Control Sample is much more evident for 
type 3+4 objects, with a strong segregation of radio sources toward red colors with respect to the Control Sample (Figure \ref{ubilate} upper panel): this difference is present, altough
with far less significance, also in the U-B colors (KS probability of 0.01).
In this case, the variable which drives this difference seems to be the redshift, as  galaxies with redshifts below
 $0.5$ have a less different distribution (KS probability of 0.01) 
than those at higher redshift  (KS probability $\sim 10^{-12}$). This is due to the fact that at higher 
redshift the Control Sample containss more blue galaxies in the B-I colors than the Complete Sample. 
Opposite behaviour is detected for the U-B color, where no difference between radio loud and parent sample is found at higher redshifts, 
while a difference with a K-S probability of $\sim 0.0007$ is present at lower redshifts.

An interesting question is whether the color difference of type 3+4 radio loud galaxies is due to some interaction between
 radio emission and optical spectrum or simply because redder galaxies tend to be more luminous in optical band  (due to the well known color-magnitude relation for galaxies) and, given a radio-optical ratio, also in the radio wavelengths.
To investigate this point, we took the Control Sample of type 3+4 galaxies (10466 objects) and  randomly assign
to each galaxy a radio-optical ratio extracted from the observed distribution (see Section 6).
With this value and given the optical luminosity of the chosen galaxy,  we computed its radio power. Then, we removed all
objects whose simulated flux is below that of the VVDS-VLA radio survey and verified that the power distribution is similar to that observed.
The resulting number of expected radio sources is 140 compared to the observed 143. Note that with this procedure we assumed the null hypothesis that the radio power is 
independent from the galaxy color but is depending only to the  optical luminosity.
As a result, the fraction of type 3+4 mock galaxies with B-I$<0.5$ is $\sim 26 \%$, significantly higher than  the observed $\sim 8\%$. 
This means that, at a given optical luminosity,  only the redder late type galaxies tend to have higher radio power, 
possibly indicating a larger amount of dust.  In other words, higher star formation rates (which make a late type  galaxy detectable in the radio bands) 
seem to occur in objects richer in dust (which makes the galaxies redder in the optical colors).
\\
%%%%%%%%%%%%%%%%%%%%%%%%%%%%%%%%%%%%%%%%%%%%%%%%%%%%%%%%%%%
\section{The Radio Luminosity Function }

Luminosity functions are a powerful tool to investigate the evolution of astronomical objects.
 However, in the case of radio sources it is more correct to refer to a bivariate luminosity function, i.e. the 
composite distribution of optical and radio luminosities. Therefore, one has to face  the possible 
physical evolution of the sources and  biases induced by the flux limits in the two bands.
This latter problem is particularly severe in the optical bands and at high redshift. As shown 
by  \cite{Ilbertbias}, in a magnitude limited sample a purely K-correction effect can induce a variation of the  population mix
of red and blue galaxies, even in the absence of evolution. Moreover, at different redshifts any single photometric
 band records different regions of the galaxy spectra making the interpretation of the results more difficult.

In order to overcome these difficulties, we choose to limit our optical sample to a
B rest frame luminosity, i.e a quantity independent from  redshift. The $M_{B}=-20$ limit
permits a galaxy to be obseved in the whole adopted redshift range eliminating the optical constraint
in the  estimation of the observable maximum volume. The luminosity function has been estimated with a standard
 $ 1/V_{max}$ method \citep{vmax} in the redshift range [0-1.1] ([0-1.0] for early type galaxies).
 The radio K-correction has been estimated
assuming that the radio spectrum is a power law (defined as $S \propto \nu^{\alpha}$) with a slope $\alpha=-0.7$.
As a check we recomputed the luminosity function estimates with the
observed 610 MHz-1.4 GHz indexes (for 310 objects within our limits) presented by \cite{Bondigmrt}: the results remain unchanged.
 We present the results without considering the completeness correction as determined in Table 3 
of \cite{Bondi}, which statistically corrects for radio sources lost at low fluxes. 
However, we have repeated the analysis also by weighting the $ 1/V_{max}$ of each radio galaxy by the 
corresponding flux--dependent correction factor: the resulting corrections have the effect of an average 
increase of the luminosity functions of $\sim 10 \%$ for logP(W Hz$^{-1}$)$<23$ in each redshift bin. 
For higher powers the corrections are larger (between $20$ and $60 \%$) but are in the same directions 
for all redshift bins and therefore our conclusions do not change.
 
\begin{figure*}
   \centering
    \includegraphics[width=\hsize]{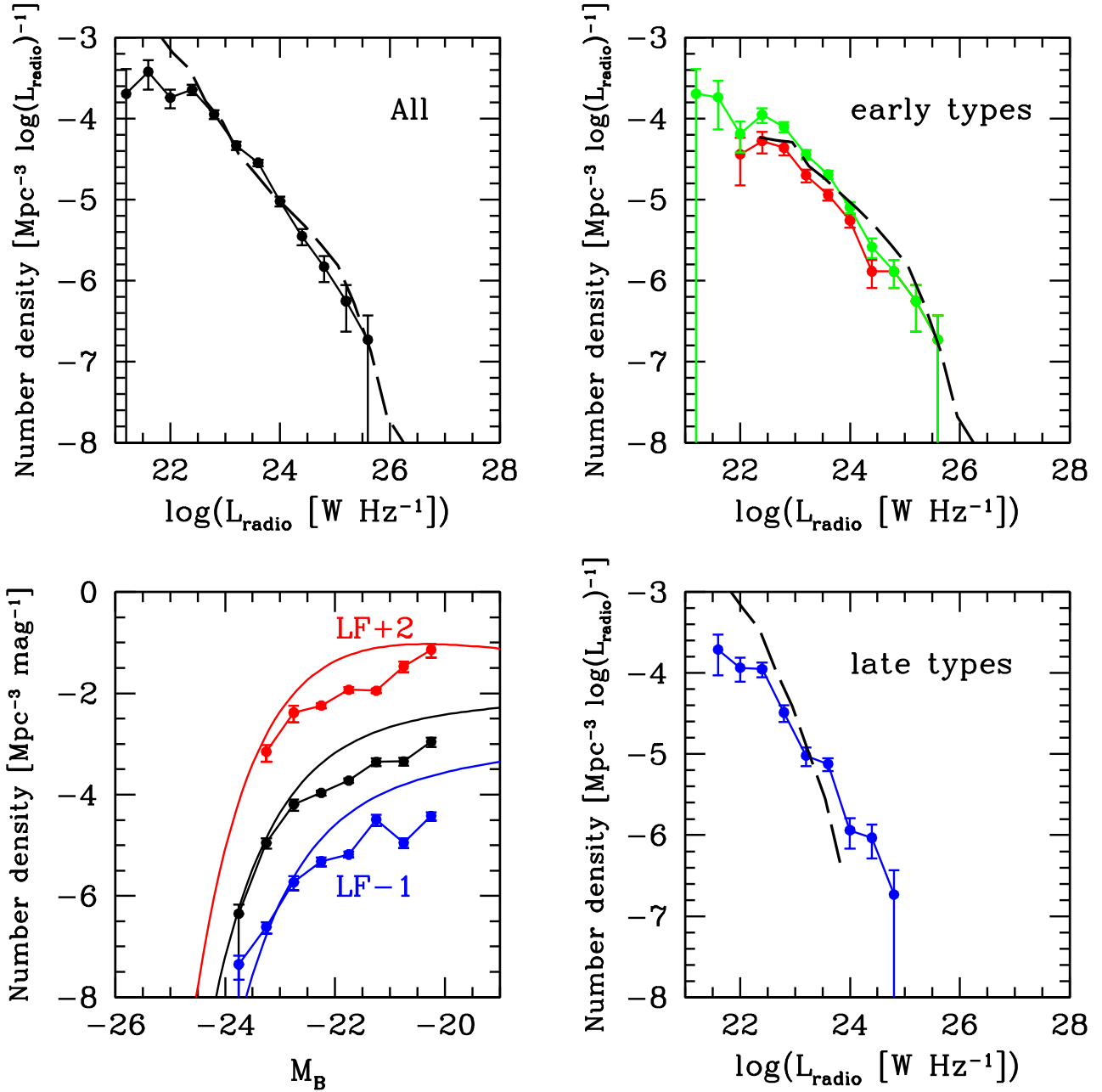}
      \caption{The radio luminosity functions of the Complete Sample for different galaxy types (points connected by solid lines) compared with NVSS/FIRST/SDSS survey 
results \citep{Best}, represented by a dashed line.
{\it Upper left panel:} all galaxies. {\it Upper right panel:} early type galaxies; in red the type 1 sample and in green
the type 1+2 sample.
{\it Lower right panel:} late type galaxies; in blue: type 3+4 sample. {\it Lower left panel:} 
Optical luminosity function of radio emitting galaxies (connected points) compared with the luminosity function of the optical galaxies in the Control Sample in the same redshift bin (solid curves).  Black: all galaxies; red: type 1 galaxies; blue: type 3+4 galaxies.
 The curves of the various samples have been vertically shifted for clarity (+2 for early type and -1 for type 1+2).}
         \label{comparison}
   \end{figure*}

In Figure \ref{comparison} the radio luminosity functions of the Complete Sample
(upper left panel), of early type (upper right panel ) and late type  (lower right panel) galaxies are compared with the corresponding 
low redshift results from  \cite{Best} (dashed lines). 
In particular, in the early type panel the red curve correponds to type 1 galaxies, while the green curve represents 
the combined type 1 + 2 sample, used to increase the statistics;  in the late type panel the blue curve represents the combined type 3+4 sample.

The Complete Sample and the early type luminosity functions are in reasonable agreement with the  \cite{Best}
estimate, also considering that their optical classification is slightly different, i.e. based on lines present in the galaxy spectra.
 The small systematic shift between our type 1 luminosity function
and the corresponding  \cite{Best} curve could be due to the different limiting optical magnitude, as the SDSS sample limited to $M_B<-19$.

In order to check the effect of different optical luminosity limits, we derived luminosity functions for the Complete Sample with two 
different absolute magnitude limits, $M_B>-19$ and  $M_B>-21$. For this test we restricted the redshift range to 
 $z<0.7$ in order to maintain the full visibility of the optical galaxies. We find that the two luminosity functions are essentially the same for 
logP(W Hz$^{-1}$)$>22.7$, with the optically bright radio luminosity function loosing $\sim 20 \% $ of objects in the range of logP(W Hz$^{-1}$) $[22-22.7]$.
 Therefore we can conclude that all differences between our estimates and those of \cite{Best} are due to the
redshift evolution of radio sources.

 The type 3+4 galaxies luminosity function is flatter than the corresponding luminosity function of star forming galaxies 
 of \cite{Best} and this remains true also considering only the most extreme population (type 4 galaxies).

Assuming a power law shape for the radio luminosity functions, we find a slope of $-1.78^{+0.07}_{-0.07}$
for type 1+2 galaxies and of $-1.84^{+0.12}_{-0.11}$ for type 3+4 objects.

In Figure \ref{comparison}, lower left panel, we show the comparison between the optical luminosity function 
of radio emitting galaxies and the corresponding luminosity function of all galaxies in the same redshift bin
\citep[obtained following][]{Zucca}.
Black curves correspond to  all galaxies, red curves correspond to type 1 objects, and blue to type 3+4 galaxies.
Connected points are the $1/V_{max}$  estimates for the radio samples, while  
solid curves are Schechter luminosity functions for optical galaxies, obtained with the 
 standard parametric STY method for estimating optical luminosity functions \citep{STY}.
For clarity, the curves for various samples have been shifted vertically, by 
a value of +2 for type 1 and -1 for type 3+4.

Working with luminosity functions means that the ratio between the two curves for a given type in this panel is directly related to the probability for a galaxy to be radio emitter.
 For luminous galaxies ($M_B<-22.5$)
the probability is higher than for fainter ones and this seems to be independent from  galaxy type. 
The probability for faint ($M_B>-21$)  objects is $\sim 7\times  10^{-2}$ and $1.7\times 10^{-2}$ for type 3+4 and type 1, respectively, and increases by more than a factor 3-4 for 
brighter galaxies.
Note the flattening of the radio luminosity functions at low powers: this effect is outlined by \cite{Cara} in estimating the 
luminosity function with the  $1/V_{max}$ in presence of evolving parent populations.

 %--------------
%--------------------------------------

   \begin{figure*}
   \centering
    \includegraphics[width=\hsize]{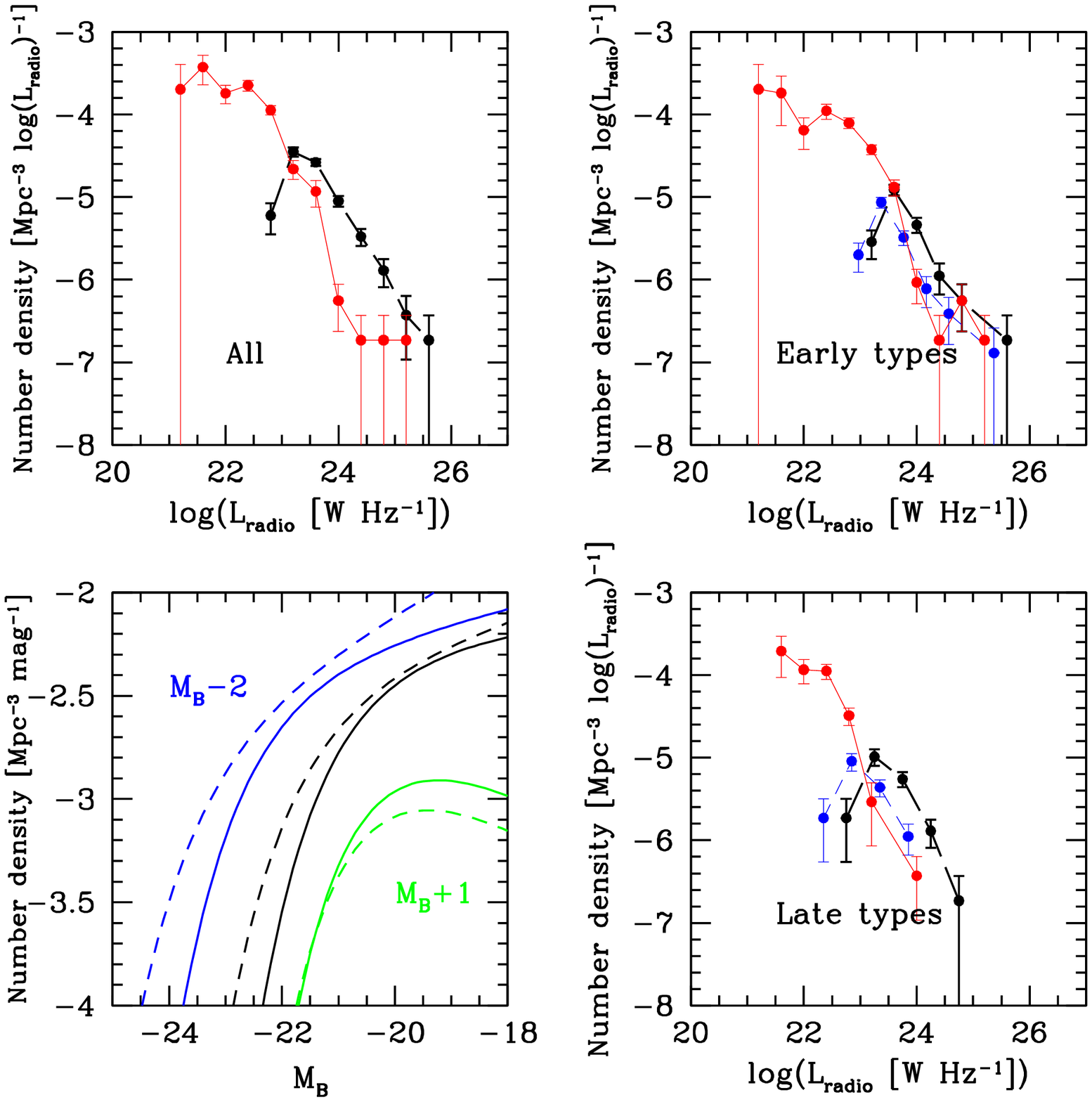}
\caption{ 
{\it Upper left panel}: Radio luminosity function of all radio emitting galaxies
with redshift below $0.5$ (red curve) and in the range [0.5-1.1] (black curve). 
{\it  Upper right panel}: the same but for type 1+2 sample and in the redshift intervals [0-0.7]
and [0.7-1.0]. Blue dashed line is the de-evolved" luminosity functions (see text).
 {\it Lower right panel}: the same but for type 3+4 sample and in the redshift intervals [0-0.5] and [0.5-1.0].
 Blue dashed line is the de-evolved" luminosity functions (see text).
{\it Lower left panel:} optical luminosity functions for the Control Sample
divided in the same types and in the same redshift intervals as previous panels. Dashed lines correspond to the high redshifts bins, while solid lines are the
lower redshifts samples. Color codes are the following: black for the Complete Sample, green for type 1+2, and blue for type 3+4 samples. For clarity the curves of the variuos samples have been shifted
 in the magnitude axis ($M_{B}+1$ for type 3+4 and $M_{B}-2$ for type 1+2
samples).}
         \label{lum_evol}
   \end{figure*}
%--------------------------------------

In Figure \ref{lum_evol} we show the evolution of the luminosity function  as a function of  redshift. In the upper left panel the black curve represents radio galaxies of our Complete Sample in the [0.5-1.1] redshift interval, while the red one corresponds to the [0-0.5] bin.  It is evident a significant evolution
in the common range of radio luminosities reaching more than a factor 10 in the
range logP(W Hz$^{-1}$) [24-25]. In the upper right and lower right panels the same is reported for type 1+2 and type 3+4 samples, respectively. We remind that given 
the different K-correction effect on absolute magnitudes, we limited the type 1+2 sample to redshift $< 1.0$.
 Moreover the limit between high and low redshift range is set to 0.7 for type 1+2 and to 0.5 for type 3+4 sample,
in order to have reasonable numbers of galaxies per bin (see Table \ref{numtab}).  
These different limits have the advantage to show the maximum difference between samples.
In order to increase the statistics, we  show the type 1+2 sample instead of only type 1 galaxies.
 Considering the most extreme early type objects the results do not change but the significance  decreases.
The evolution of early type galaxies is mild, althought present, and it corresponds to a density increase of a factor $\sim 5$ in the logP(W Hz$^{-1}$) range [23.5-24], while 
the evolution of type 3+4 sources is stronger. This strong evolution is consistent with that of
\cite{Afonso05}, which was found at  $z<0.5$, and \cite{Smolcic} for the deeper VLA-COSMOS sample.  
Assuming an evolving law for luminosity and density evolution, of the form of $(1+z)^{\beta}$ and  $(1+z)^{k}$ respectively,
we found that the best fit is $\beta=2.76$ and $k=0.04$ for type1+2 and $\beta=2.63$ and $k=0.43$ for type 3+4
objects. These values are consistent with the literature  \citep{Brown,Haarsma,Hopkins}.
However, the errors are quite large and the $\beta$ and $k$ parameters are strongly correlated: 
our aim is only to show that the early type galaxies present luminosity evolution with small density evolution, while late type galaxies show both evolutions. 
These results, obtained using a sample with a large redshift range and with high statistics,
 are consistent with the very early estimates of \cite{Colla} and \cite{Auriemma}.

The detected evolution could have mainly two origins: a) the evolution of the corresponding optical luminosity functions, i.e. one has more objects and/or more optically luminous ones, which would translate in more luminous radio sources in the case of
a constant radio-optical ratio; b) the evolution with redshift of the radio optical ratios;
or a combination of both a) and b).

In order to give an idea of the behaviour of the optical luminosity functions, which enter in point a),
 we show in  Figure \ref{lum_evol} (lower left panel) the STY estimates \citep[computed following ][]{Zucca}, for all galaxies (black), type 1+2
(green) and type 3+4 (blue) samples  in the same redshift bins used for the radio samples.
The curves have been shifted for clarity in the magnitude axis by +1 for type 1+2 and by -2 for the type 3+4 class.

The main difference which can be seen in this panel is a 
brightening by one magnitude with a small steepening of the slope from low to high redhifts
for type 3+4 objects. For type 1+2 objects the main difference is that at high redshifts the galaxies are 0.2 mag brighter (from the best fit of $M^*$ in the Schechter function) 
and $\sim 30\%$ less numerous than today. 
These results are consistent with those published for the VVDS survey  \citep[see ][Table 3]{Zucca}.

In the next section we will estimate the second parameter entering in the radio luminosity function evolution, i.e. the radio-optical ratio
(which regards point b). 

We ``de-evolved" the high redshift radio luminosity functions taking
into account our conclusion regarding the optical 
luminosity function and the radio-optical ratio for the two classes.
 This exercise aims to verify whether the detected evolution of the bivariate 
radio-optical luminosity function is the mirror of the evolution of both the optical luminosity 
function and the radio-optical ratio.

In order to ``de-evolve'' the bivariate luminosity function of type 3+4 sample, we cut the higher redshift bin at $M_{B}=-21$, assuming an average one-magnitude evolution
of the  corresponding optical function (limited at $M_B<-20$ at $z=0$) and no evolution of the radio-optical ratio.
 The luminosity function cut eliminates all optical galaxies entered at high redshifts 
in the radio sample and therefore responsible of the radio density evolution.

For type 1+2 galaxies we corrected for density and luminosity evolution  as detected for the optical luminosity function
and a variation of $\sim 1.4$ in the radio-optical ratio (see Section 6).
 
The resulting "de-evolved" luminosity functions are presented as blue points in the right panels of Figure \ref{lum_evol}. The remaining discrepancy between 
these functions and the low redshift luminosity functions is likely due to contamination of BLAGN and NLAGN (see Section 2.1).
 
Therefore, we estimated whether the difference between the "de-evolved" high redshift and the low redshift radio luminosity function of type 3+4 objects  is consistent 
with the expected AGN contamination, assuming that all AGN are in the high redshift bin. 
The number excess of the "de-evolved" luminosity function with respect to the low redshift one is 12, consistent again with the expected contamination of $15\pm 5$ (see Section 2.1).
%--------------------------------------
\section{The Radio-Optical ratio}

In the previous section we showed the radio luminosity distribution for a sample with an optical absolute magnitude limit.  
A related way to show the relationship between the radio and optical emission is through the radio-optical ratio.
Classically, it was estimated by using the quantity $S_{(mJy)} \times 10^{(m-12.5)/2.5}$, where $S$ is the radio flux and {\it m} 
is the apparent magnitude of a radio galaxy. In our case, this formula is not satisfactory because it does not take into account the
shift between the observed and the rest frame fluxes, which is important in our sample because of its large range in 
redshift. For this reason we compute the ratio between the radio luminosity ($L_R$) and the optical B band luminosity as
$$L_{R}/L_B = L_R /[5.08\times 10^{0.4(5.48-M_{B})} \times 10^{29}]$$
\\
which represents a rest--frame radio-optical ratio, with both $L_R$ and $L_B$ in  $erg\ s^{-1}$.
However, this is not enough to estimate the correct radio-optical ratio distribution. In fact, the cut in radio power 
as a function of redshift 
induced by the flux 
limited survey  introduces a Malmquist bias in the ratio distribution, while this problem is not present
for the optical data. For this reason, a given radio galaxy with an optical luminosity $L_B '$ and radio luminosity $L_R'$ 
is visible  to its maximum redshift $z_{max}'$ determined by its radio luminosity and the limiting flux of the survey: 
the corresponding radio-optical ratio is therefore observable only within the 
corresponding $V_{max}'$.
If one wants to compare the occurrence of various radio-optical ratios within a fixed volume $V$, the  radio-optical ratio
of each single radio galaxy has to be weighted with the factor $V / V_{max} $. Moreover, comparing frequencies means that
the histograms must be divided by the expected $\sum V / V_{max}$ and not simply by the number of radio sources.

 %--------------------------------------
   \begin{figure}
   \centering
 \includegraphics[width=\hsize]{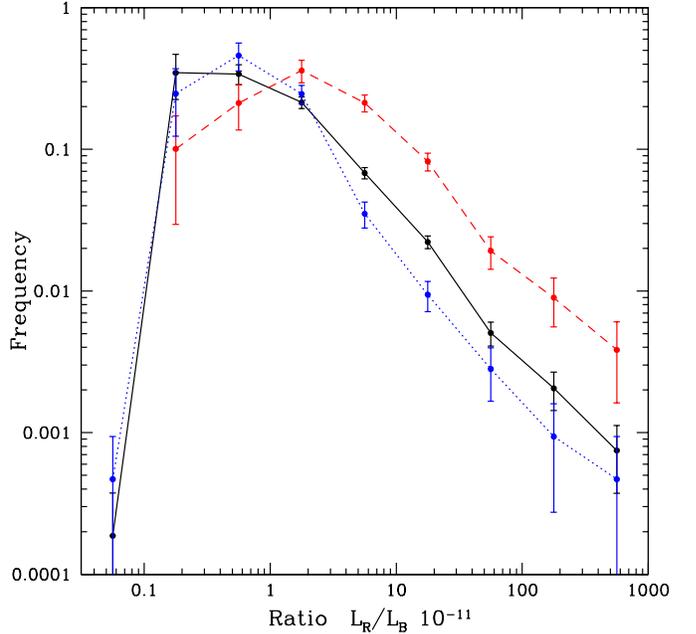}
      \caption{Frequency of the radio-optical Ratio in our sample. Solid line is the Complete Sample, dashed line corresponds to
 type 1+2 galaxies and dotted line to type 3+4 galaxies}
         \label{Lradloptratio}
   \end{figure}
%--------------------------------------
   \begin{figure}
   \centering
\includegraphics[width=\hsize]{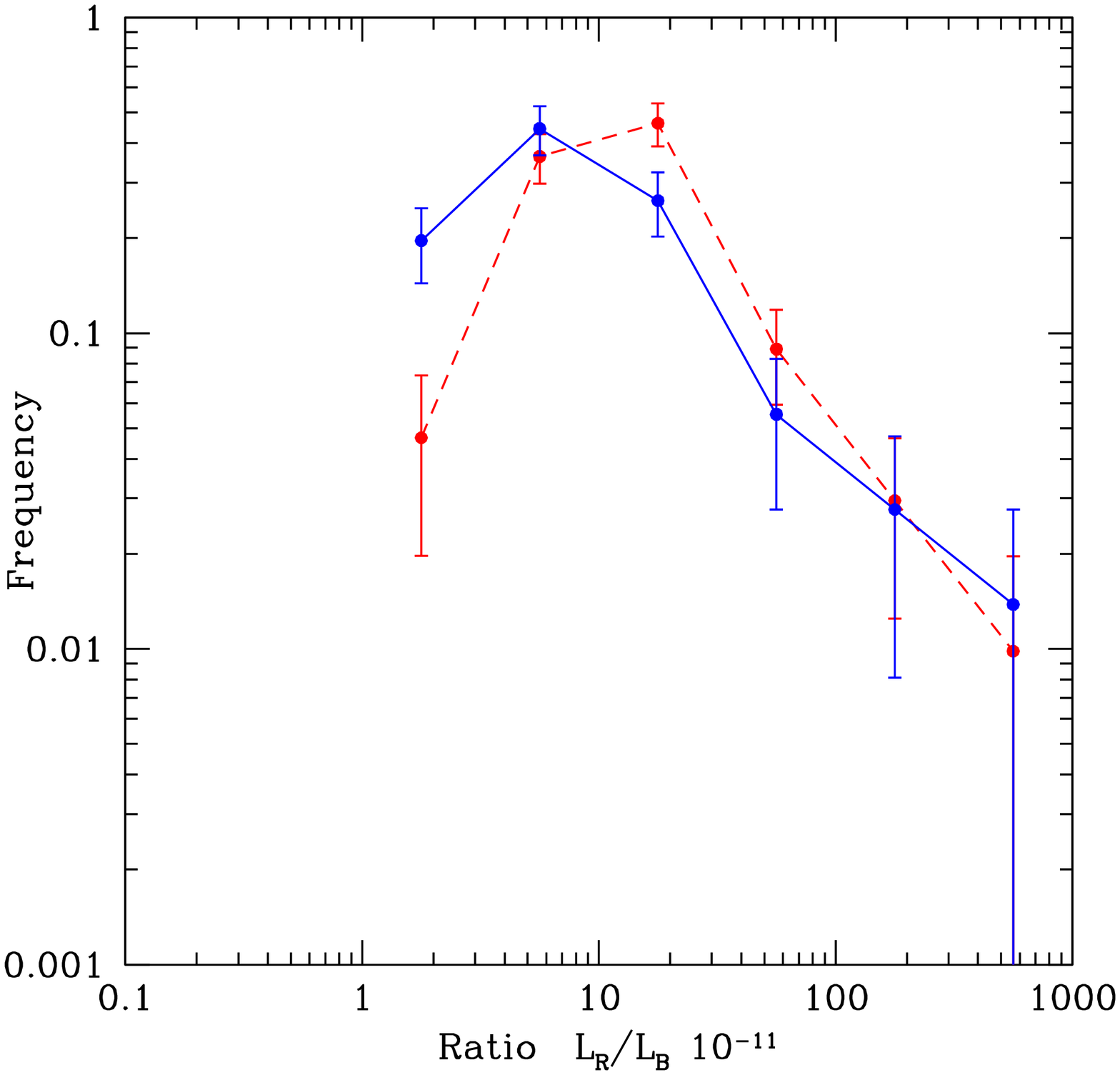}
\includegraphics[width=\hsize]{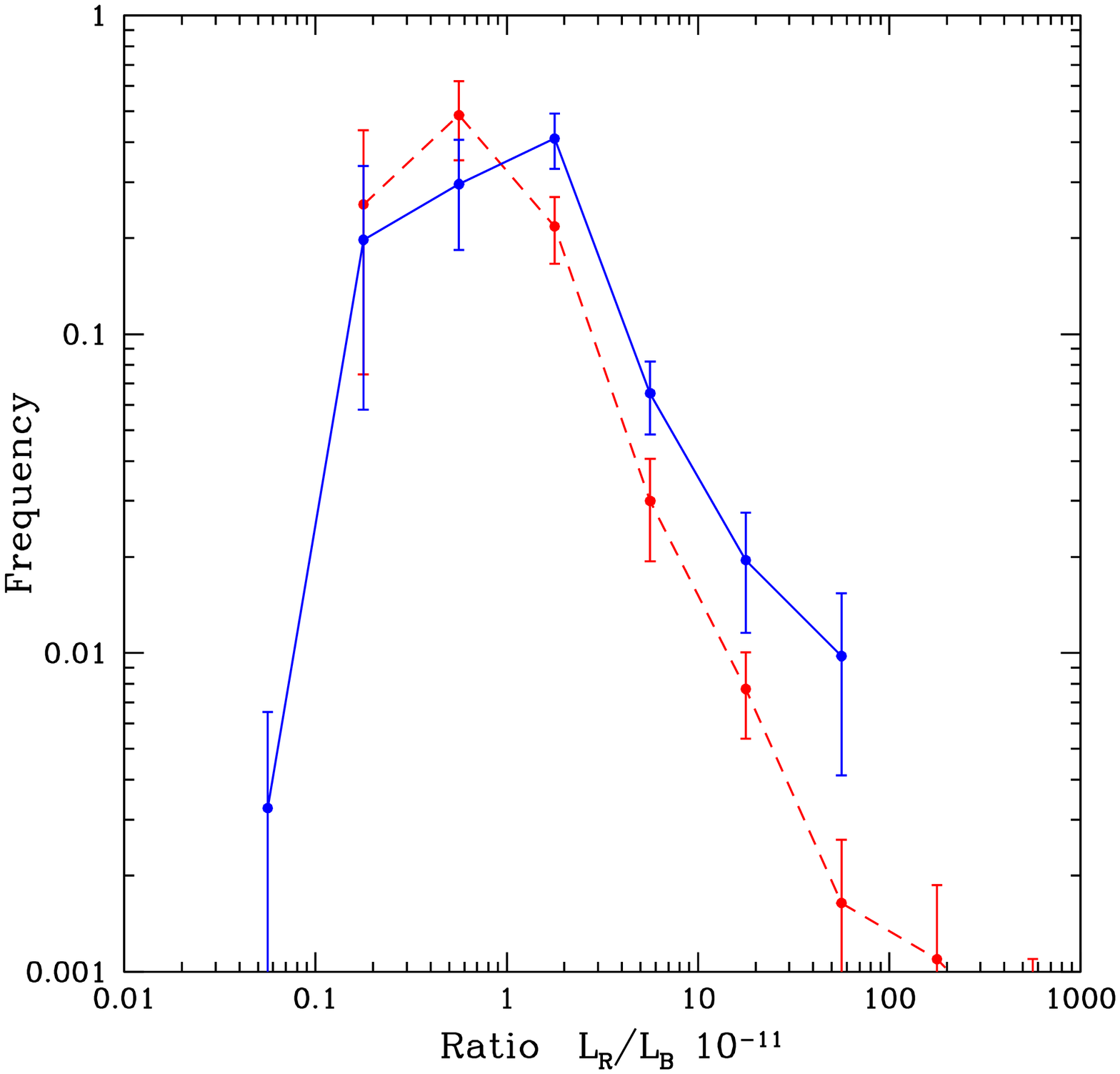}
\caption{{\it Upper panel:} frequency of radio-optical ratio as a function of redshift for type 1+2 galaxies. 
Solid line represents low redshift ($z<0.7$) galaxies, while dasehd line refers to high redshift galaxies.
 To be comparable, the samples have been limited at logP(W Hz$^{-1}$)$>$23.1
{\it Lower panel:} the radio-optical ratio for type 3+4 galaxies is plotted for objects with $M_B<-21.5$
(red dashed line) and $M_B>-21.5$ (blue solid line).} 
   \label{Radloptratio}
   \end{figure}
%
%--------------------------------------
%--------------------------------------
The radio-optical distributions derived in this way for the Complete Sample
and type 1+2 and type 3+4 samples are shown in Figure \ref{Lradloptratio}.
 The distributions are significantly broadened around the mean. For late type galaxies 
this is likely due to different dust content (seen as distribution in blue luminosity) corresponding to a given radio power,
which is directly related to the star formation rate. For early type
objects this is a measure of the AGN power variance, presumably related to the accretion
efficiency, at a given optical luminosity.

The radio-optical distribution of early type galaxies is shifted towards higher values
with respect to that of late type objects.
In order to study the evolution of the radio-optical ratio in the same redshift bins of the luminosity function, it is 
necessary to consider only those radio luminosity ranges 
which are always observable in the considered redshift range. For this reason, we cut the type 1+2 sample at logP(W Hz$^{-1}$)$>23.1$ and 
the type 3+4 at logP(W Hz$^{-1}$)$>22.8$, because the two classes have different high redshift limits.  
Although the number statistics are significantly reduced by these cuts, it is possible to say that no significant evolution is
present for type 3+4 objects.
This result is not surprising because if the shape of the initial mass function remains constant with  redshift, the number of 
resulting radio emitting supernova remnants depends 
only on the total number of stars  formed. 
However, we detected a slight dependence of the radio-optical ratio as a function of the optical magnitude, 
with the radio-optical distribution presenting a tail toward higher ratios for galaxies with $M_B<-21.5$ 
(Figure \ref{Radloptratio}, lower panel). 
 We checked whether the increase of the radio-optical ratio for bright 
galaxies is due to the presence of an AGN using the upper 
panel of Figure \ref{lumtosfr} (see Section 7). The star formation values
for bright galaxies are consistent with the logSFR(radio)- logSFR(optical) relation.
 
For type 1+2 objects (Figure \ref{Radloptratio}, upper panel) the situation appears to be more complicated. 
It seems that there is a shift of a factor $\sim 1.4$ between the low redshift and high redshift bins. 
This would imply that the efficiency to feed the central black hole was a bit higher
in the past. We used the reported factor 1.4 for deriving the "de-evolved" luminosity function described in Section 5.

\section{Star Formation Rate and Stellar Mass }

\begin{figure}
   \centering
 \includegraphics[width=\hsize]{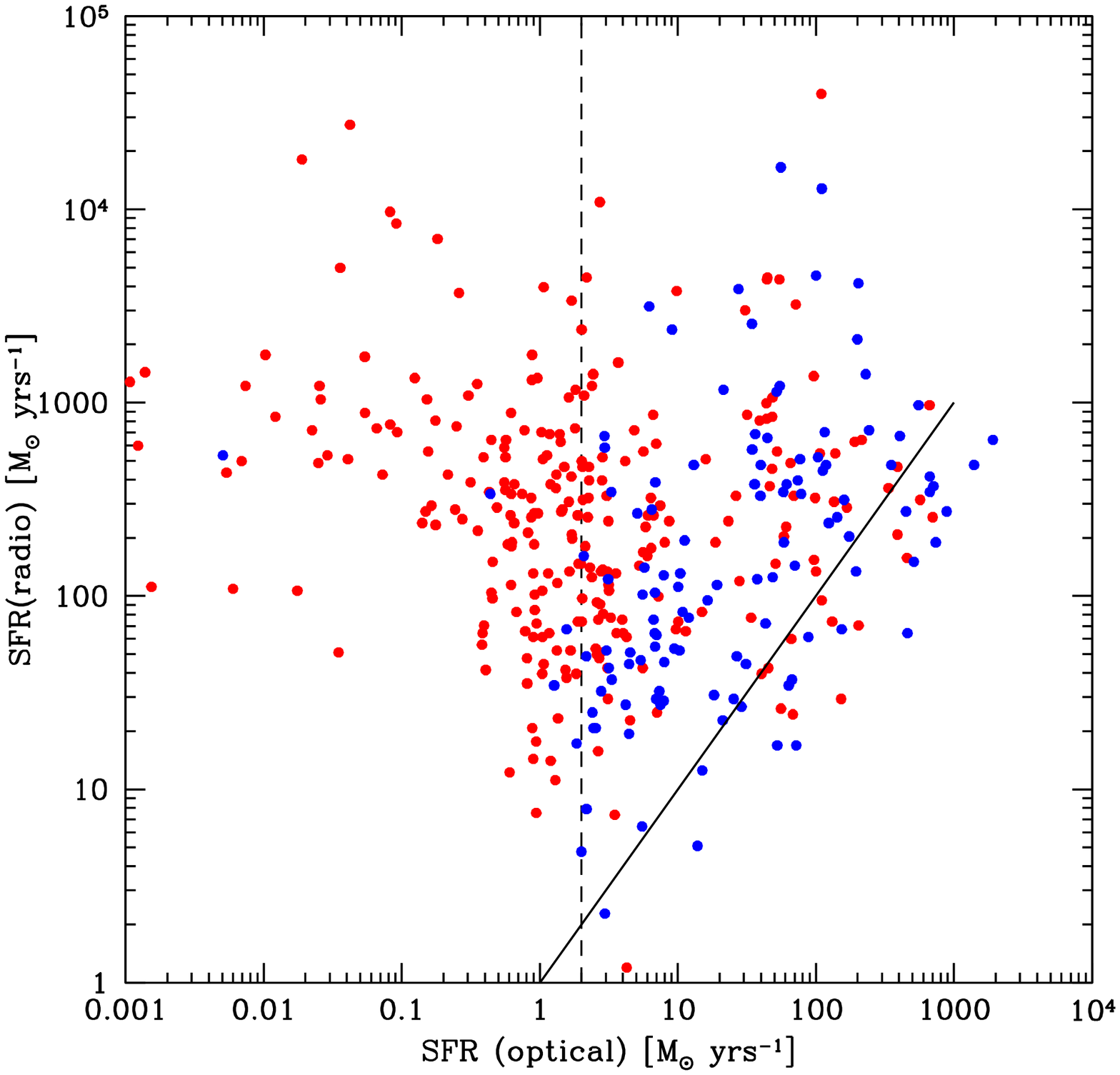}
\includegraphics[width=\hsize]{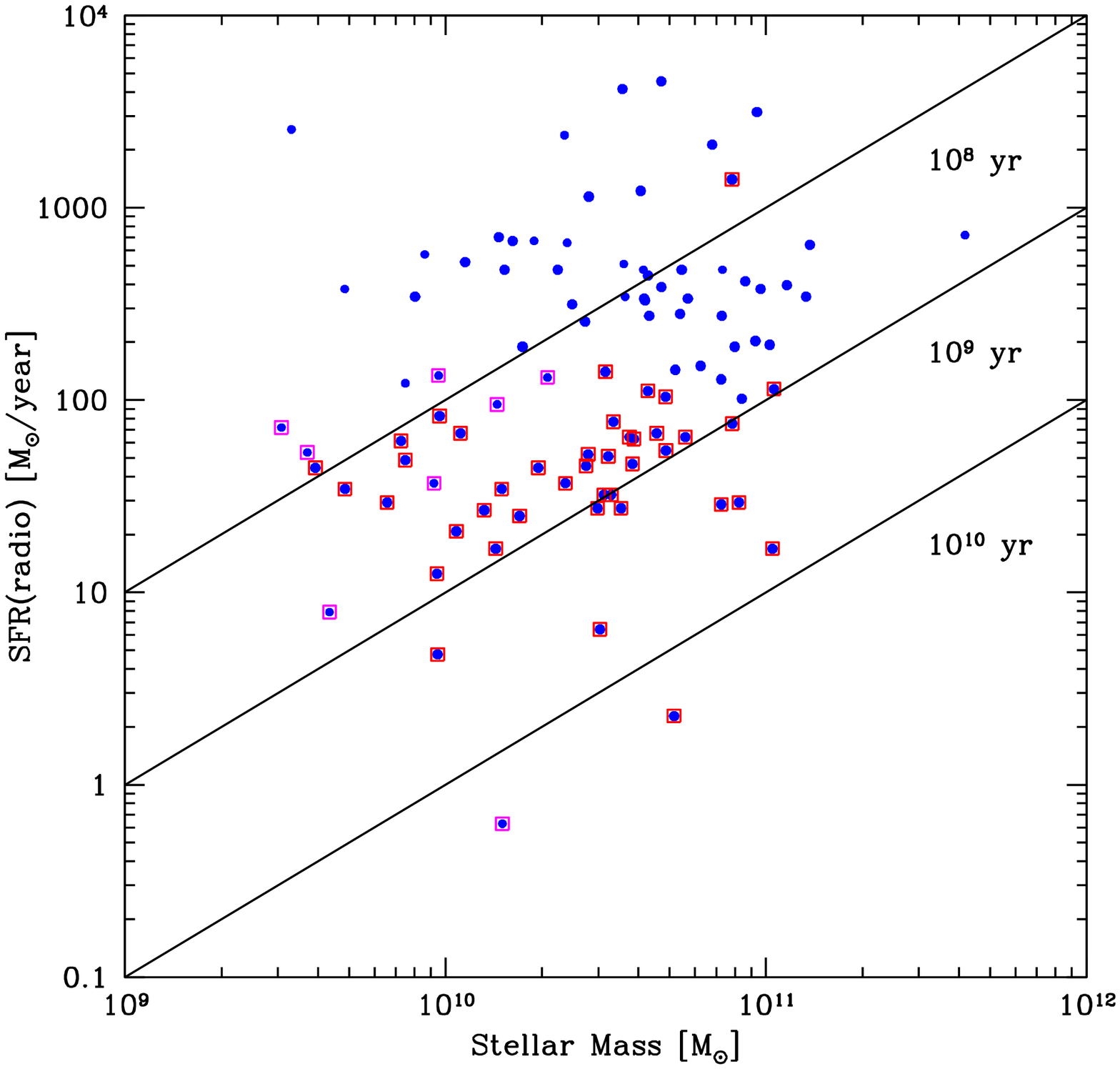}
     \caption{{\it Upper Panel:} Star formation rate computed from the optical versus the star formation rate determined from the radio 
emission. Red and blue  points correspond to type 1+2 and 3+4 objects, respectively. The vertical dashed line corresponds to the value of $2
M_{\odot} yrs^{-1}$, while the solid line corresponds to logSFR(optical)=logSFR(radio) and is plotted for reference.
 {\it Lower Panel:}  Stellar masses versus star formation rate derived from the radio emission for type 3+4 galaxies; squares 
represent the low redshift ($z<0.5$) objects.}
         \label{lumtosfr}
   \end{figure}

Star formation rates (SFR) and  stellar masses for galaxies have been computed from the photometric quantities,
following the method described in detail in \cite{Pozzetti}, by matching the data to a set of templates computed with various 
star formation histories and dust absorption.  The star formation rates and the stellar masses have been calculated 
using the \cite{Chabrier} initial mass function, while the equations used in the literature to compute SFR from the radio power
use the \cite{Salpeter} initial mass function. This implies that the SFR from the radio and optical band could be slightly different.
For consistency, we converted our quantities by estimating the median difference of 
log(SFR$_{Salpeter}$) - log(SFR$_{Chabrier}$) $\sim$ 0.186, 
when the two SFR are compared. In order to be consistent with other optical papers, when we compare
SFR and stellar mass distributions of radio loud galaxies and of the Control Sample we used the original 
\cite{Chabrier} initial mass function estimates.
In the upper panel of  Figure \ref{lumtosfr} we show the star formation rate  computed for the Complete Sample following this procedure
based on the optical spectral distribution, compared with the star formation rate estimated following the formula of \cite{Haarsma}
($ SFR(radio)=L_{1.4 GHz} / 8.36 10^{20}  M_{\odot}\ yr^{-1} $) , which uses only radio powers at $1.4$ GHz \citep[see][]{Condon}.
In the literature there are other formulae to estimate the star formation rate with different assumptions, as for example that of
\cite{Bell} and \cite{Cram}, which give rates a factor 2.16 and 4.35 lower than  that of \cite{Haarsma}, respectively.
 
The majority of early type galaxies ($\sim 70 \%$) are segregated in the region with  SFR(optical)$<2 M_{\odot}\ yr^{-1}$:
this fraction increases to  $\sim 82 \%$ when considering only the extreme type 1 objects.
 On the other hand,  almost all ($ \sim 86\%$) late type galaxies  have star formation rates larger than $2 M_{\odot}\ yr^{-1}$.
This is not surprising because both the star formation rate and the type classification depend on the difference between red and
 blue-UV colors (mainly driven by the 4000 $\AA $ break). The early type galaxies with  SFR$>2 M_{\odot}\ yr^{-1}$
show similar $M_B$ with respect to the low star formation rate counterparts but a brighter ultraviolet color.
This UV excess, which is responsible of the high star formation rate,  has little, if any, impact onto the classification.

There are some weak trends in the SFR(radio)-SFR(optical) plot for the two galaxy classes, type 1+2 showing an anticorrelation, while type 3+4 seem to be correlated.
 To evaluate the significance of the correlations we applied the Spearman rank test finding the values of $\rho_{1+2}=-0.173$ and  $\rho_{3+4}=0.474$ with a probability of the null 
hypothesis (no-correlation) of $0.002$ in both cases.   
However, we assumed that the radio emission for type 1+2 objects is AGN induced and therefore this relationship is to be regarded as 
radio power versus star formation rate.
It is worth to note that the type 3+4 objects are not symmetrically distributed around the relation logSFR(radio)=logSFR(optical), 
being only $\sim 19 \%$ of the points below the equality line. 
This result can be explained by the increasing role of dust 
with increasing star formation rates. Note that the factor between radio power and star formation rate which maximize the fraction of 
objects with  logSFR(radio)$<$logSFR(optical) is that of \cite{Cram}, leading to $34 \%$.

 The lower panel of  Figure \ref{lumtosfr} is analogous to Figure 3 of \cite{Cram}. The diagonal lines correspond to the characteristic time scales 
$\tau$= 10$^{8}$, 10$^{9}$ and 10$^{10}$ yr defined by Cram et al. as 
$M_{\odot}/SFR_{1.4GHz}$. \cite{Cram} claimed that there is a tendency for galaxies that have already formed many stars to support a higher current star formation; 
this effect is present also in  our sample if we ignore the effect of different redshift. Dividing the sample in two redshift bins (above and below 0.5),  
we find that at the same stellar mass the star formation rate as estimated from the radio band is higher at higher redshifts. We checked this result also taking all radio galaxies 
with logP(W Hz$^{-1}$)$>22.8$ in order to avoid biases 
due to different power ranges in different redshift ranges, finding the same result, i.e. there is no trend for the star formation rate 
with stellar mass, but the mean value of the star formation distribution was higher in the past.
 This means that
the specific star formation rate is decreasing  with both redshift and with stellar mass \citep[see][]{Vergani}.

\begin{figure}
   \centering
 \includegraphics[width=\hsize]{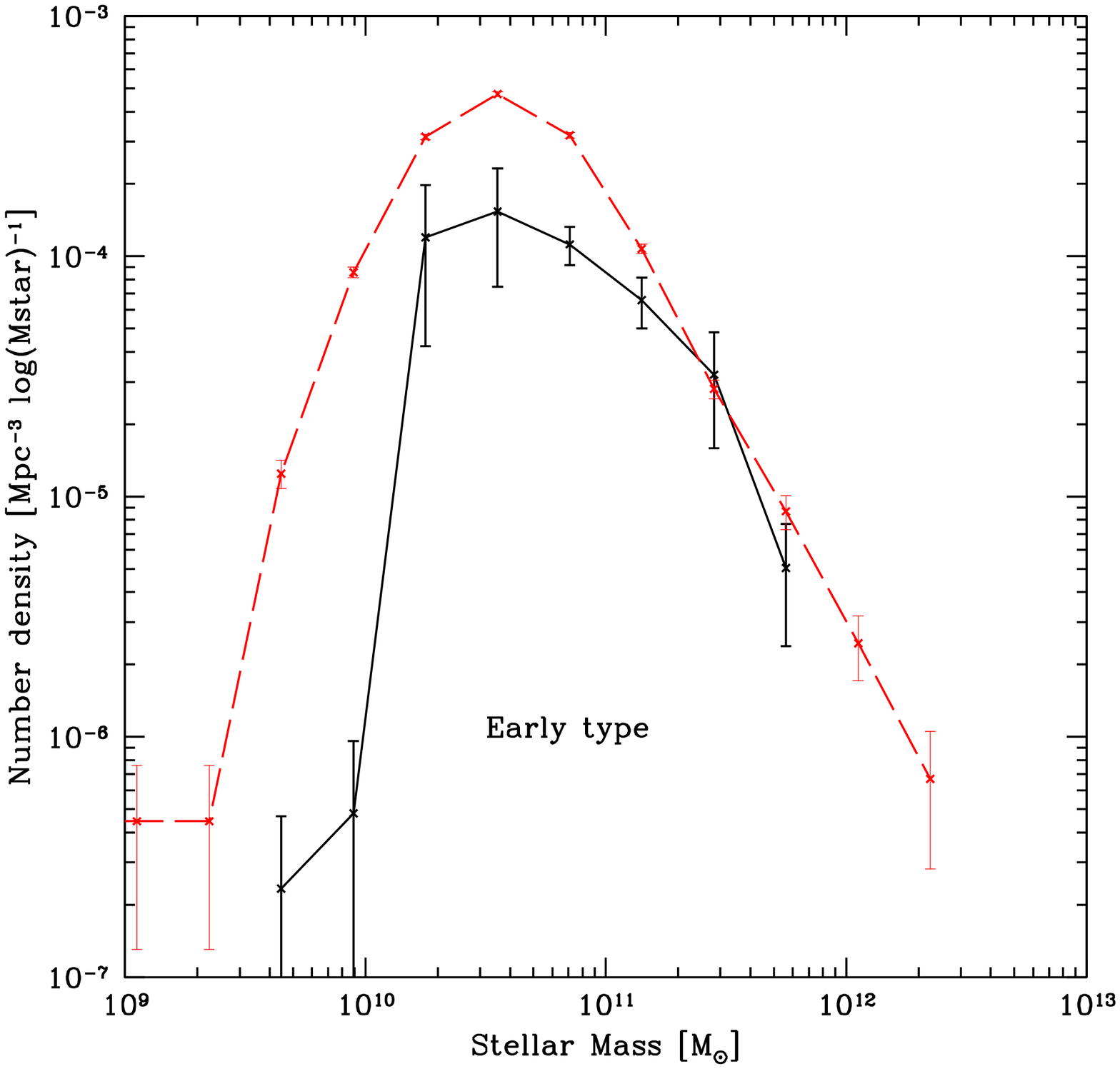}
\includegraphics[width=\hsize]{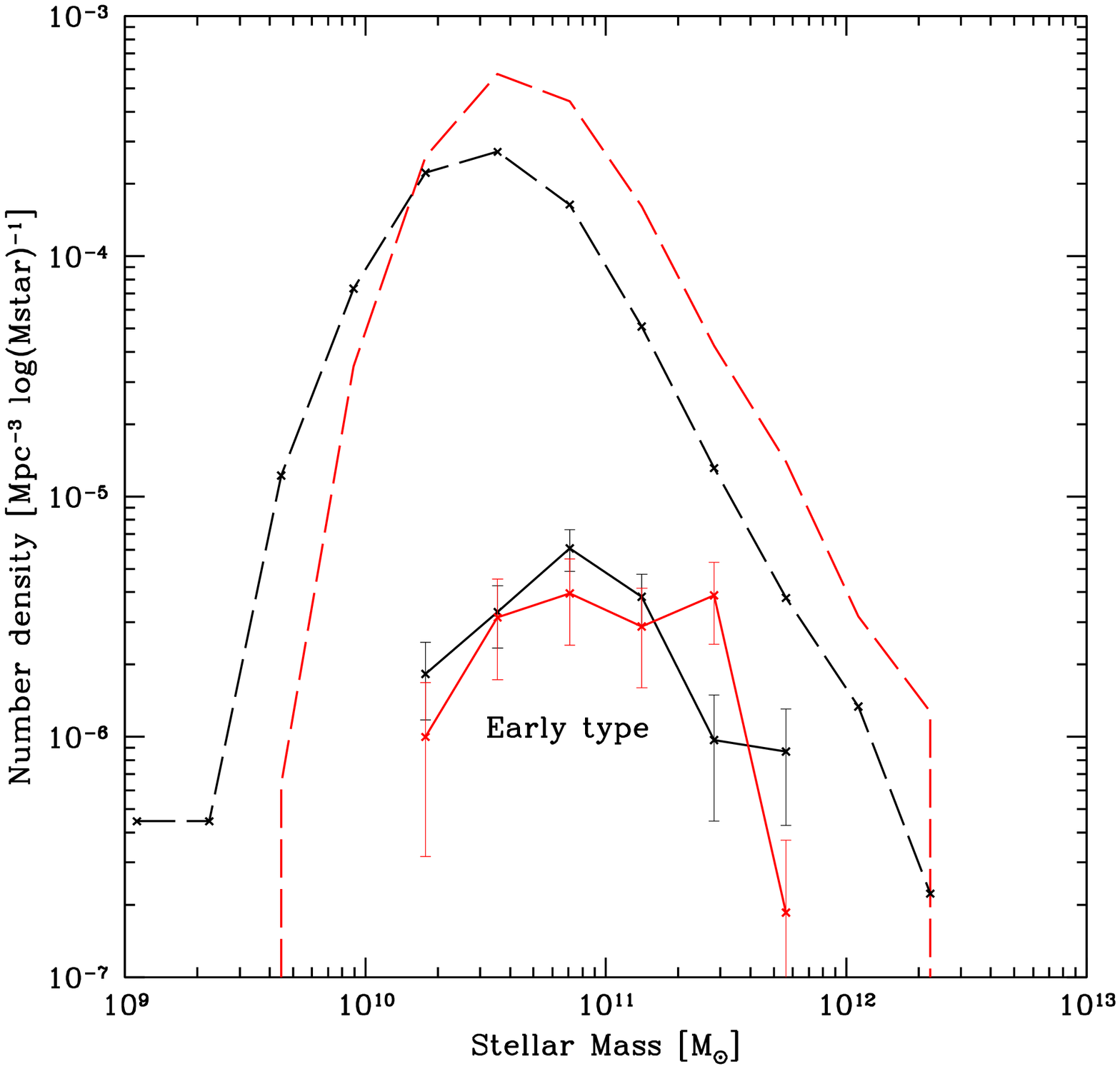}
   \caption{{\it Upper panel:} Stellar mass functions of the type 1+2 radio sample (black solid line) and the type 1+2 control sample
 (red dashed curve).
 {\it Lower panel:} in red the low redshift  ([0-0.7]) subsample for type 1+2 galaxies and in black the high redshift sample ([0.7-1.0]). Dashed curves correspond to the control samples, solid curves to radio loud objects.}
       \label{mstar}
   \end{figure}

\begin{figure}
   \centering
\includegraphics[width=\hsize]{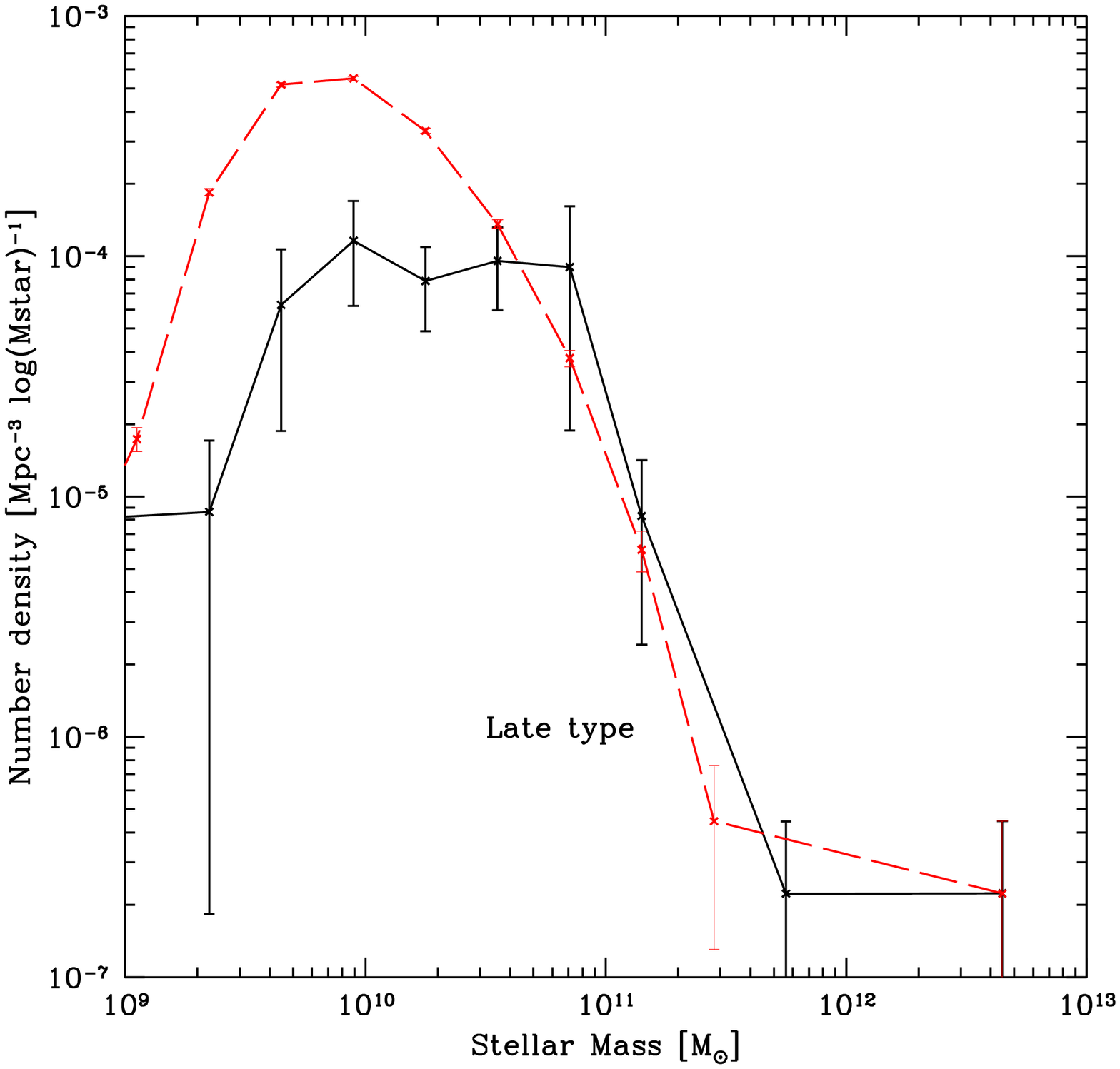}
 \includegraphics[width=\hsize]{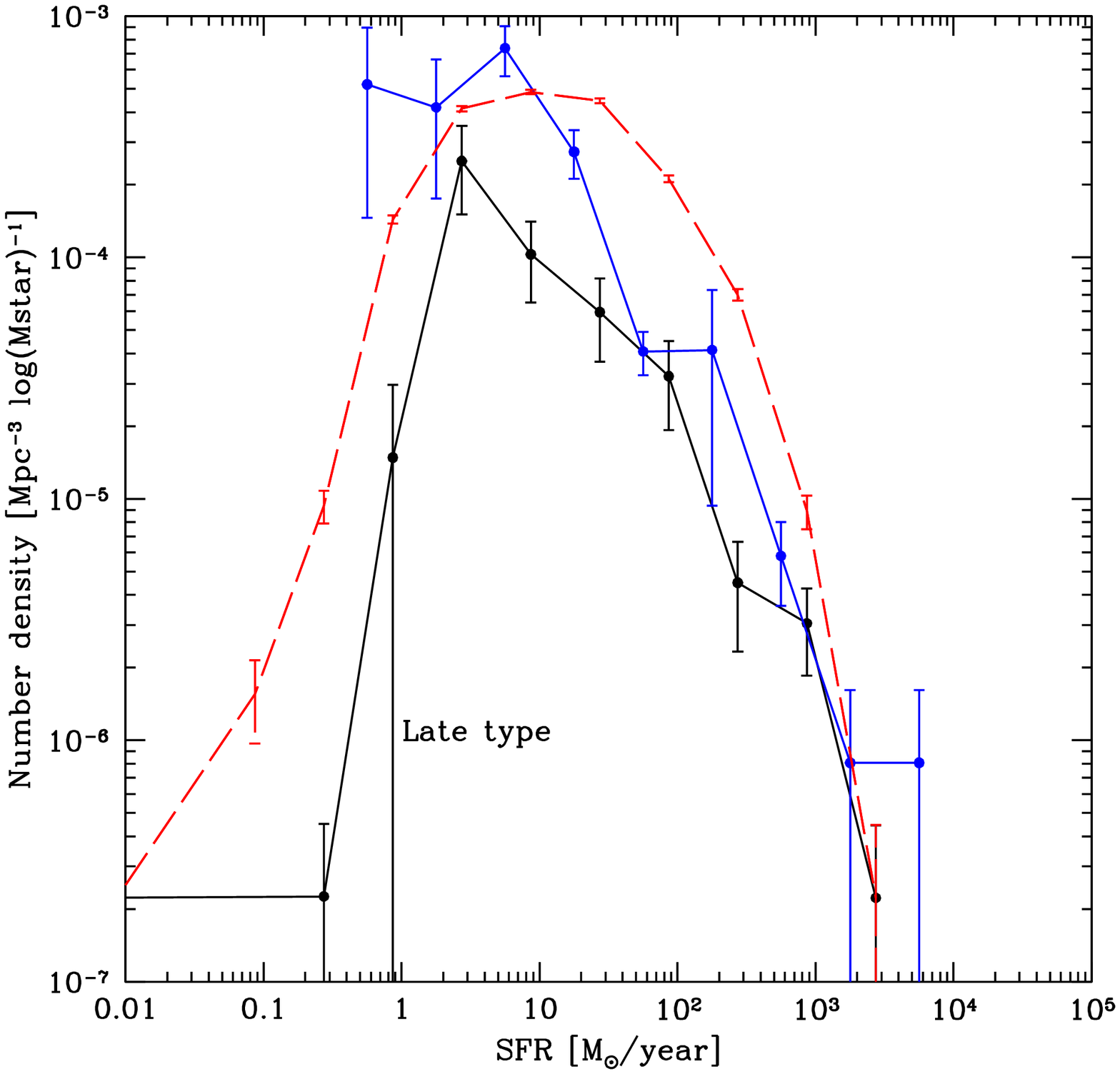}
\caption{{\it Upper panel:} Stellar mass functions  of the type 3+4 radio Sample (black solid line) and the type 3+4 Control Sample 
(red dashed line).
 {\it Lower panel:} Star formation rates distribution of the radio rample (red solid line) and the Control Sample (black dashed line) for type 3+4 galaxies. The blue line corresponds to star formation rate computed from the radio emission.}
         \label{sfr}
   \end{figure}

As done for the radio-optical ratio, in Figure \ref{mstar} and \ref{sfr} we plot the stellar masses and the star formation rate 
distribution weighted by $V_{max}$: 
correcting stellar masses with the maximum volumes is equal to estimate the mass function.  
Our mass functions are not fitted by a Schechter function because the effect of increasing mass incompleteness with increasing
redshift is important in our large redshift ranges. However, we are interested in the comparison between radio and Control Sample
and we assumed that the mass incompleteness is not a function of the radio properties and the same argument holds also for the 
star formation.

Similarly to what has been found for  the optical luminosity function of the radio emitting galaxies,
also for stellar masses of type 1+2 objects the fraction of radio loud objects increases with the mass 
and almost all galaxies with  $M> 10^{11}$ M$_\odot$ are radio loud (upper panel of Figure \ref{mstar}).

In the lower panel of Figure \ref{mstar} we estimated the stellar mass functions in the two redshift bins we used for the luminosity functions. It is clear that the number density of radio sources
is not changing  with cosmic time, although the overall number of massive galaxies is increasing
\citep[see][]{Pozzetti}. This probably means that the AGN is already active at high redshift and no new activity formed in between.
For what concerns the star formation rate, the distributions of radio loud and radio quiet type 1+2 objects are similar 
for SFR$>1\ M_\odot $ yr$^{-1}$.  The two curves are only shifted by a factor less than $2$.

The distributions for late type galaxies are significantly different, where the SFR of radio emitting objects is 
always lower than that of the Control Sample, although the difference is decreasing with increasing star formation. 
This is true using the star formation rates estimated in the optical bands. When we use
the star formation rate based on the radio power, we obtain the blue line of Figure \ref{sfr}, 
more similar to the distribution of star formation rates of the Control Sample.
 Note that the behaviour shown for late type galaxies in Figure \ref{sfr} is similar 
to the relations found by \cite{Seymour} (see their figure 4).

%%%%%%%%%%%%%%%%
%%%%%%%%%
%%%%%%%%%%
%%@@@@@@@@@@@@@@@@@@@@@@@@@@@2

\section{Star formation density evolution}

Assuming that radio emission from late type galaxies is entirely due to star formation, the estimation of the star formation density evolution follows in a natural way.
The most direct and used method is the determination of the radio luminosity function and its integral in different redshift bins. 
The integral of the radio luminosity function is directly related to the actual star formation rate.
The drawbacks of this method are
essentially the poor statistics (few hundreds of radio objects at most are observed 
within the entire range of redshifts) and the fact that at different redshifts the sampled galaxies are different, being optically more luminous at higher redshifts. 

Another way to approach the problem is that of considering the optical sample 
as a tracer of radio emission. 
In this case the star formation density could be written as
$$  SFR_{density} = C \int \phi(L') \left( \int R D(R) L' dR \right) dL'   $$ 
where $R$ and $D(R)$ are the radio-optical ratio and its probability distribution  (derived as radio-optical distribution
in Section 6), $L$ is the blue absolute luminosity  (see Section 2) , $\phi(L)$ is the optical luminosity function  and $C$ is the factor 
converting the radio luminosity density to star formation density   \citep[assumed to be that of][see Section 7]{Haarsma}.

%%%%%%%%%%%%%%%%
   \begin{figure}
   \centering
   \includegraphics[width=\hsize]{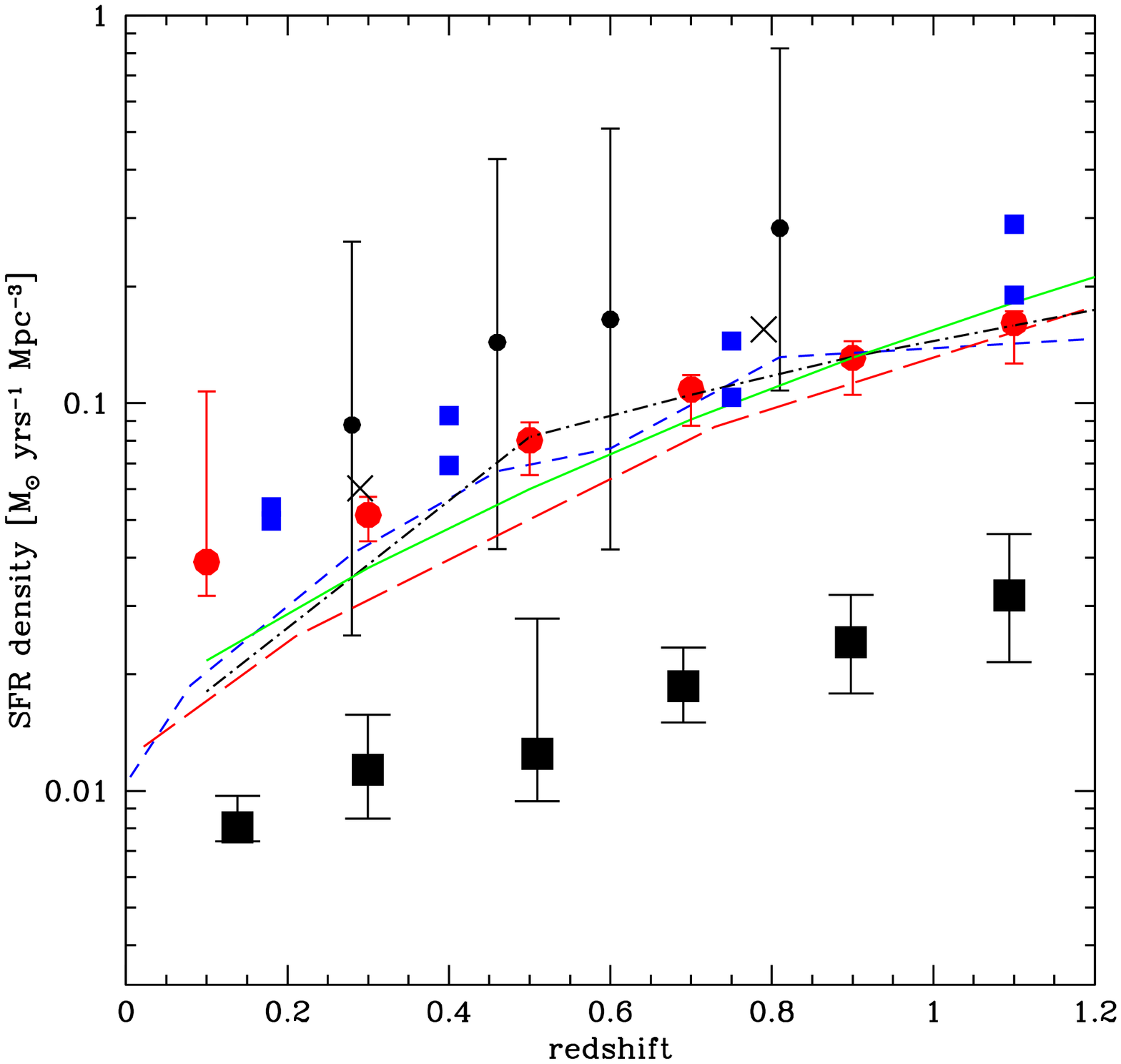}
      \caption{Star formation rate density as a function of the reddshift. The red points are our estimates, the black points those of 
\cite{Haarsma} as corrected by   \cite{Hopkins}, the green solid line is  the fit
of \cite{Hopkins}, the black dot-dashed line is from \cite{Cram}.
The blue squares are the points of  \cite{Smolcic} and crosses are the estimates of \cite{Seymour}. Large black squares with error bars correspond to the VVDS dust uncorrected values, while red and blue dashed lines
correspond to the star formation rate density estimated from near-IR and H$\alpha$ data as reported by \cite{Tresse}.}
         \label{sfr_density}
   \end{figure}
%%%%%%%%%%%%%%%

Note that all the quantities in this equation have been estimated above. In particular, we take into account the difference between the optical luminosity function of the Control Sample 
and that of radio emitting galaxies and the dependance of the radio-optical ratio on the optical magnitude.

We refitted the optical luminosity function in the B-band on the same sample and using the same method of \cite{Zucca} 
but considering together type 3 and type 4 galaxies. For $\phi(L)$ 
we used the fitted Schechter function, which depends on three parameters ($\alpha$, $M^*$ and $\phi^*$).
Errors are estimated considering the variance in the fit of the optical luminosity function
and considering the poissonian error on the radio-optical distibution function. 
In fact, we computed various Schechter functions on the border of the one sigma error ellipses and considered the 
maximum and minimum functions as  part of the error induced by the optical luminosity function. 
Similarly, we have taken as upper and lower limit for the radio-optical distribution function  the one sigma fluctuations in the histogram of Figure \ref{Lradloptratio} and again we assumed  the maximum and minimum of the computed star formation density as error induced by the radio-optical distribution. 
Finally, we added in quadrature these values with the previous ones.

 For the last two redshift bins, where the slope $\alpha$ of the optical luminosity function  is not constrained, we fixed the value of the slope to $\alpha=-1.33$, which is the fitted 
value of the two bins considered together.
 Therefore, for the last two bins the errors given by our procedure are underestimated. 
 The values are reported in Table \ref{sfrtab}.
In Figure \ref{sfr_density} the estimated star formation density is compared with values found in the literature both from radio and other bands. The radio data has been converted to the formula 
of \cite{Haarsma}.

There is an overall consistency between all the estimators, even if our star formation density 
in the lowest redshift bin is higher than the other estimates. However,
considering that the real error bars could become a bit larger by taking into account other sources of statistical variation 
than error in the optical luminosity function fit alone, our point at $z=0.1$ could be regarded as still consistent
 with other estimators.  
Moreover, also the first bin of \cite{Smolcic},
whose data in term of limiting radio flux are similar to ours, is high.    
By comparing our values with the VVDS dust uncorrected values found by \cite{Tresse} using ultraviolet data, we find that at all redshifts there is an approximately constant shift 
 of $\Delta$ logSFR$_{density} = 0.73\pm 0.02$. There is also a weak indication of a $\sim 10 \%$ increase of this correction
with redshift.
Note that the value is very similar to the correction of $\sim 0.74$
of \cite{Hopkins01} (see equation 7)  required for the FUV  $1500 \AA$ \citep[for which the][SFR densities are computed]{Tresse} 
and a star formation rate of logSFR $\sim 1$.

Moreover, our star formation density  is fitted by an evolution of 
$(1+z)^{2.15}$, consistent with the ultraviolet light density evolution
of \cite{Tresse}. 

What is important here is the conclusion that the optical light could be used as tracer of the radio emission at high redshifts, provided that 
no significant variation of the radio-optical ratio is present. Note that we checked this point because our radio sample cover, even if with small statistics, a large fraction of the range over which the star formation history has been estimated. 

%_____________________________________________________________
%
\begin{table*}
\centering          
\label{table:2}     
\caption{Parameters of the optical luminosity function and value of the star formation rate. The symbol $^*$ means that errors are underestimated.}  
\begin{tabular}{c c c c c }     % 7 columns 
\hline\hline       
                      % To combine 4 columns into a single one 
$z$ & $M^* $ & $\alpha$ & $\phi^*$ (Mpc$^{-3})$ & log(SFR$_{density}$) (M$_{\odot}$ yr$^{-1}$ Mpc$^{-3}$)    \\ 
\hline                    
\smallskip
   0.1 & -19.81$^{+0.54}_{-1.00}$     & -1.13$^{+0.10}_{-0.10}$ &  0.0066$^{+0.0030}_{-0.0028}$  & -1.410$^{+0.440}_{-0.087}$ \\ \smallskip
   0.3 & -20.17$^{+0.19}_{-0.22}$     & -1.18$^{+0.07}_{-0.07}$  &  0.0057$^{+0.0012}_{-0.0011}$ & -1.288$^{+0.047}_{-0.068}$ \\ \smallskip
    0.5 & -21.05$^{+0.19}_{-0.22}$    & -1.29$^{+0.07}_{-0.07}$  &  0.0027$^{+0.0006}_{-0.0006}$ & -1.096$^{+0.047}_{-0.089}$ \\ \smallskip
   0.7 &  -21.23$^{+0.16}_{-0.18}$    & -1.33$^{+0.08}_{-0.08}$  &  0.0027$^{+0.0006}_{-0.0005}$ & -0.965$^{+0.038}_{-0.093}$ \\ \smallskip
   0.9 &  -21.19$^{+0.17}_{-0.18}$    & -1.26$^{+0.11}_{-0.11}$  &  0.0036$^{+0.0008}_{-0.0007}$ & -0.883$^{+0.043}_{-0.096}$ \\ \smallskip
   1.1 & -21.52$^{+0.09}_{-0.09}$     & -1.33 Fixed  &  0.0027$^{+0.0001}_{-0.0001}$ & -0.793$^{+0.020*}_{-0.104*}$ \\ \smallskip
   1.3 & -21.54$^{+0.10}_{-0.10}$     & -1.33 Fixed  &  0.0026$^{+0.0001}_{-0.0001}$ & -0.792$^{+0.003*}_{-0.106*}$ \\ 
\hline                  
\end{tabular}
\label{sfrtab}
\end{table*}
%
%_____________________________________________________________
%                                          Table with foonotes 

%%%%%%
\section{Conclusions}

In this paper we analyzed the rest frame properties of radio loud galaxies of the  VVDS-VLA Deep Field survey.
In order to avoid all difficulties intrinsic to samples containing both optical and radio limits, we 
have chosen to a) work using rest-frame quantities;  b) limit the sample to an absolute magnitude  $M_{B}<-20$
and c) consider only objects with $z<1.1$ ($<1.0$ in the case of early type galaxies). 
The two last points correspond to having an optical volume limited sample, i.e. no loss of optical galaxies with
redshift. Therefore, the only limit which remains is the radio flux limit ($\sim 80 \mu Jy$).
In this way we can control the optical properties of radio galaxies, although with paying the price of smaller statistics
 and loosing all optically faint objects.  

Following  \cite{Zucca}, we divided the galaxies in four spectrophotometric classes corresponding to early type (types 1 and 2) and late type
(type 3 and 4)  galaxies.  Furthermore, we assumed that early type galaxies have  AGN--induced radio emission and late type galaxies have star--formation induced emission.
The aim of this work was to investigate the
possible evolution of  radio galaxies knowing {\it a priori} the optical behaviour of all galaxies with our cuts in the entire CFHT-LS/VVDS survey
(taken as Control Sample).

From the redshift distribution of radio sources we detected an overdensity corresponding to two interacting clusters at  redshift  $z \sim 0.6$,
while the other redshift peaks appear to be associated with a  large scale structure (superclusters). It has been impossible to investigate further 
this point because of the relatively small number of spectroscopic redshifts and the smoothing effect due to the use of photometric redshifts.
The angular two points correlation function of radio loud galaxies does not show significant differences with respect to that of optical galaxies, 
suggesting that these objects are not living in the most extreme environments. 
By estimating the distribution of blue luminosities, stellar masses and star formation rates, it results that
the probability for a galaxy to be a radio emitter increases significantly at high values of these parameters. 

Early type (type 1) radio loud galaxies show some differences to respect to their radio quiet counterparts in the colors distribution. In particular, radio faint 
galaxies show a peak in the B-I color at B-I$\sim 1.45$, while radio bright galaxies show no difference with the Control Sample.
This class of radio sources shows evolution in the bivariate radio-optical luminosity function, mainly due to luminosity evolution.
Studying the  radio-optical ratio distribution we show that for redshifts above $0.7$ at fixed blue luminosity the galaxies were more radio luminous.
This effect is somehow compensated in the bivariate luminosity function because of the simultaneous decrease in the type 1+2 optical luminosity function.
 This luminosity evolution is likey due to a different efficiency to feed the central black hole 
and not to an increased contribution by star formation in the radio band. 
This could be seen also by considering the results of Figure \ref{lumtosfr} (upper panel) 
divided in redshift bins: the fraction of galaxies with star formation rates (as computed in the optical
band) lower than $2 M_{\odot}\ yr^{-1}$ is the same at high and low redshifts. 
This means that, unless dust absorption and star formation exactly compensate each other, 
no significantly increased star formation for this galaxy type is present. 

Type 3 and type 4 radio loud objects are significantly redder with respect to the control population.
These galaxies show a significant evolution in the bivariate luminosity function both in luminosity and density:
this behaviour can be explained if we take into account the one-magnitude evolution of this class of objects detected in the optical, without invoking large variations in the radio emission
 properties (like the radio-optical ratio).
In fact, the only dependence of the radio-optical ratio is, in this case, based on the optical luminosity.
Using the estimated radio-optical ratio we conclude that the difference in colors between radio loud and radio quiet objects is physical, and not induced by statistical effects: 
we speculated that radio loud star forming galaxies contain  on average more dust.

The star formation rate as computed from radio power and from optical colors shows a significant correlation similar to the relation betwen radio and UV-based SFR 
shown in  \cite{Cram}. By plotting the star formation rate computed from radio luminosities and the stellar mass, we show 
that at higher redshifts the population has higher instantaneous star formation at a fixed stellar mass. In other words, the specific star formation is 
decreasing with redshift.

With the knowledge of the parameters for the radio loud type 3 and type 4 galaxies, we show that it is possible to use
optical galaxies as tracers of the radio emission at various redshifts. This permits to derive an estimate of
the star formation history, which resulted to be consistent with the results of other bands and methods. 

In summary, from our data, the AGN induced radio emission is increasing with redshift because the host galaxies increased their radio-optical ratios, with the 
AGN already in place at high redshifts. On the contrary, the evolution of late type radio sources is a direct consequence of the strong evolution seen the in 
optical band and is due to the evolution of the star formation rate with redshift.

\begin{acknowledgements}
S.B. thanks C. Gruppioni, I. Prandoni, F. Pozzi for useful discussions.
This research has been developed within the framework of the VVDS
consortium.\\
The authors thank the referee for the careful reading of the manuscript and for
the comments, which improved the paper.
This work has been partially supported by the
CNRS-INSU and its Programme National de Cosmologie (France),
and by Italian Ministry (MIUR) grants
COFIN2000 (MM02037133) and COFIN2003 (num.2003020150).\\
The VLT-VIMOS observations have been carried out on guaranteed
time (GTO) allocated by the European Southern Observatory (ESO)
to the VIRMOS consortium, under a contractual agreement between the
Centre National de la Recherche Scientifique of France, heading
a consortium of French and Italian institutes, and ESO,
to design, manufacture and test the VIMOS instrument.
Based on observations obtained with MegaPrime/MegaCam, a joint  
project of CFHT and CEA/DAPNIA, at the Canada-France-Hawaii Telescope  
(CFHT) which is operated by the National Research Council (NRC) of  
Canada, the Institut National des Science de l'Univers of the Centre  
National de la Recherche Scientifique (CNRS) of France, and the  
University of Hawaii. This work is based in part on data products  
produced at TERAPIX and the Canadian Astronomy Data Centre as part of  
the Canada-France-Hawaii Telescope Legacy Survey, a collaborative  
project of NRC and CNRS.

\end{acknowledgements}

%--


\begin{thebibliography}{}
\bibitem[Afonso et al., 2005] {Afonso05}               Afonso, J., Georgakakis, A., Almeida, C., et al. 2005, \apj, 624, 135
\bibitem[Afonso et al., 2006] {Afonso06}               Afonso, J., Mobasher, B., Koekemoer A., Norris, R.P., Cram, L. 2006, \aj, 131, 1230
\bibitem[Auriemma et al., 1977] {Auriemma}             Auriemma, C., Perola, G.C., Ekers, R., et al. 1977 \aap, 57, 41
\bibitem[Barger et al., 2007] {Barger07}               Barger, A.J., Cowie, L.L., \& Wang, W.H. 2007, \apj, 654, 764
\bibitem[Bardelli et al., 2009] {Bardelli}             Bardelli, S., et al., 2009, \aap, in preparation 
%\bibitem[Barthel, 2006] {Barthel}                     Barthel, P.D. 2006, \aap, 458, 107
\bibitem[Bell, 2003] {Bell}                            Bell, E.F. 2003, \apj, 586, 794
\bibitem[Benn et al., 1993] {Benn}                     Benn, C.R., Rowan-Robinson, M., McMahon, R.G., Broadhurst, T.J., Lawrence, A. 1993, \mnras, 263, 98    
\bibitem[Best et al., 2005] {Best}                     Best, P.N., Kauffmann, G., Heckman, T.M., Ivezic, Z. 2005, \mnras, 362, 9
\bibitem[Bondi et al., 2003] {Bondi}                   Bondi, M., Ciliegi, P., Zamorani, G., et al. 2003, \aap, 403, 857
\bibitem[Bondi et al., 2007] {Bondigmrt}               Bondi, M., Ciliegi, P., Venturi, T., et al. 2007, \aap, 463, 519
\bibitem[Bongiorno et al., 2007] {Bongiorno}           Bongiorno, A., Zamorani, G., Gavignaud, I., et al. 2007, \aap, 472, 443
\bibitem[Brown et al., 2001] {Brown}                   Brown, M.J., Webster, R.L., Boyle, B.J. 2001, \aj, 121, 2381 
\bibitem[Bruzual \& Charlot, 1993] {bc93}              Bruzual, G., Charlot, S. 1993, \apj, 405, 538
\bibitem[Cara \& Lister, 2008] {Cara}                  Cara, M., Lister, M.L. 2008 \apj,  686, 148
\bibitem[Chabrier, 2003] {Chabrier}                    Chabrier, G. 2003, \pasp 115, 763  
\bibitem[Ciliegi et al., 2005] {Ciliegi1}              Ciliegi, P., Zamorani, G., Bondi, M., et al. 2005, \aap, 441, 879
\bibitem[Coleman et al., 1980] {CWW}                   Coleman, G.D., Wu, C.C., \& Weedman, D.W. 1980, \apjs, 43, 393
\bibitem[Colla et al., 1975] {Colla}                   Colla, G., Fanti, C., Fanti, R., et al. 1975, \aap, 38, 209 
\bibitem[Condon, 1992] {Condon}                        Condon J.J 1992, \araa,  30, 575  
\bibitem[Cram et al., 1998] {Cram}                     Cram, L., Hopkins, A., Mobasher, B., Rowan-Robinson, M. 1998, \apj, 507, 155
\bibitem[Franceschini et al., 1988] {Franceschini}     Franceschini, A., Danese, L., Toffolatti, L., de Zotti, G. 1988, \mnras, 233, 157
\bibitem[Franzetti et al., 2007] {Franzetti}           Franzetti, P., Scodeggio, M., Garilli, B., et al. 2007, \aap, 465, 711
\bibitem[Gavignaud et al., 2006] {Gavignaud}           Gavignaud, I., Bongiorno, A., Paltani, S., et al. 2006, \aap, 457, 79
\bibitem[Giacintucci et al., 2004] {Giacintucci}       Giacintucci, S., Venturi, T., Bardelli, S., Dallacasa, D., Zucca, E. 2004, \aap, 419, 71
\bibitem[Gruppioni et al., 1999] {Gruppioni99}         Gruppioni, C., Mignoli, M., \& Zamorani, G. 1999, \mnras, 305, 297
\bibitem[Gruppioni et al., 2003] {Gruppioni}           Gruppioni, C., Pozzi, F., Zamorani, G., et al. 2003, \mnras, 341, L1
\bibitem[Haarsma et al., 2000] {Haarsma}               Haarsma, D.B., Partridge, R.B., Windhorst, R.A., Richards, E.A. 2000, \apj, 544, 641
\bibitem[Hammer et al., 1995] {Hammer}                 Hammer, F., Crampton, D., Lilly, S.J., Le F\`evre, O., Kenet, T. 1995, \mnras, 276, 1085 
\bibitem[Hopkins et al., 2001]{Hopkins01}              Hopkins, A.M., Connoly A.J, Haarsma D.B., Cram L.E. 2001,  \aj, 122, 288
\bibitem[Hopkins, 2004] {Hopkins}                      Hopkins, A.M. 2004, \apj, 615, 209
\bibitem[Ilbert et al., 2004] {Ilbertbias}             Ilbert, O., Tresse, L., Arnouts, S., et al. 2004, \mnras, 351, 541
\bibitem[Ilbert et al., 2005] {Ilbertlf}               Ilbert, O., Tresse, L., Zucca, E., et al. 2005, \aap, 439, 863
\bibitem[Ilbert et al., 2006] {Ilbertphz}              Ilbert, O., Arnout, S., McCracken, H.J., et al. 2006, \aap, 457, 841
%\bibitem[Iovino et al., 2005] {IovinoJK}               Iovino, A., McCracken, H.J., Garilli, B., et al. 2005, \aap, 442, 423
\bibitem[Ivison et al., 2007] {Ivison}                 Ivison, R.J., Chapman, S.C., Faber, S.M., et al., 2007, \apj, 660, L77
%\bibitem[Kauffmann et al., 2003] {Kauffmann}           Kauffmann, G., Heckman, T.M., Tremonti, C. et al. 2003, \mnras, 346, 1055 
%\bibitem[Kinney et al., 1996] {Kinney}                 Kinney, A.L., Calzetti, D., Bohlin, R.C. et al. 1996, \apj, 467, 38
\bibitem[Le F\`evre et al., 2005] {Lefevre2005a}       Le F\`evre, O., Vettolani, G., Garilli, B., et al. 2005, \aap, 439, 845
\bibitem[Lilly et al., 2007] {zcosmos}                 Lilly, S.J., Le F\`evre, O., Renzini, A., et al. 2007, \apjs,  172, 70 
\bibitem[Lonsdale et al., 2003] {swire}                Lonsdale, C.J., Smith, H.E., Rowan Robinson M., et al. 2003, \pasp, 115, 897
%\bibitem[Mahabal et al., 1999] {Mahabal}               Mahabal, A., Kembhavi, A., \& McCarthy, P.J. 1999, \apj, 516, L61
\bibitem[Magliocchetti et al., 2002] {Magliocchetti}   Magliocchetti, M., Maddox, S.J., Jackson, C.A., et al. 2002, \mnras, 333, 100
\bibitem[McCracken et al., 2003] {McCrackenVVDS}       McCracken, H.J., Radovich, M., Bertin, E., et al. 2003, \aap, 410, 17
\bibitem[McCracken et al., 2008] {McCrackenLegacy}     McCracken, H.J. et al. 2008, \aap, in preparation
\bibitem[McCracken et al., 2008] {McCracken}           McCracken, H.J., Ilbert, O., Mellier Y., el al., 2008, \aap, 479, 321 
\bibitem[Owen et al., 1999] {Owen99}                   Owen, F.M., Ledlow, M.J., Keel, W.C., Morrison, G.E. 1999, \aj, 118, 633
\bibitem[Paltani et al., 2008] {Paltani}               Paltani, S., et al. 2008, \aap, in preparation 
\bibitem[Pozzetti et al., 2007] {Pozzetti}             Pozzetti, L., Bolzonella, M., Lamairelle, F., et al. 2007, \aap, 474, 443
\bibitem[Prandoni et al., 2001] {Prandoni}             Prandoni, I., Gregorini, L., Parma, P., et al. 2001, \aap, 369, 787
\bibitem[Salpeter, 1955] {Salpeter}                    Salpeter E.E.  1955, \apj, 121, 161
%\bibitem[Tadhunter et al., 2002] {Tadhunter}          Tadhunter, C., Dickinson, R., Morganti, R. et al. 2002, \mnras, 330, 977
\bibitem[Sandage, Tammann \& Yahil 1979] {STY}          Sandage, A., Tammann G.A., Yahil, A. 1979, \apj, 232, 352
\bibitem[Schmidt, 1968] {vmax}                         Schmidt, M. 1968, \apj, 151, 393
\bibitem[Seymour et al., 2008] {Seymour}               Seymour, N., Dwelly, T., Moss, D., et al. 2008, \mnras 368, 1695 
\bibitem[Smolcic et al., 2008] {Smolcic}               Smolcic, V., Schinnerer, E., Zamorani, G., et al. 2008, \apj, in press (arXiv:astro-ph/0808.0493)
\bibitem[Tresse et al., 2007] {Tresse}                 Tresse, L., Ilbert, O., Zucca, E. et al. 2007, \aap, 472, 403
\bibitem[Vergani et al., 2008] {Vergani}               Vergani, D., et al. 2008, \aap, in preparation 
\bibitem[Warren et al., 2007] {UKIDSS}                 Warren, S. J., Cross, N. J. G., Dye, S., et al. 2007, arXiv:astro-ph/0703037
\bibitem[Windhorst et al., 1985] {Windhorst}           Windhorst, R.A., Miley, G.K., Owen, F., et al. 1985, \apj, 289, 494
\bibitem[Zucca et al., 2006] {Zucca}                   Zucca, E., Ilbert, O., Bardelli, S., et al. 2006, \aap, 455, 879 
\end{thebibliography}
\end{document}